\documentclass[a4p, 12pt]{article}%%%%
\usepackage{setspace}%%%%
\usepackage{latexsym}%%
\usepackage{amssymb}%% 
\usepackage{amsmath} %
\usepackage{amsthm} %
\usepackage{amscd} %
\usepackage{mathrsfs} %%\mathscr{S}}%
\newtheorem{thm}{Theorem}[section]
\newtheorem{coro}[thm]{Corollary}
\newtheorem{lem}[thm]{Lemma}
\newtheorem{prop}[thm]{Proposition}
\theoremstyle{definition}
\newtheorem{defn}[thm]{Definition}
\theoremstyle{remark}
\newtheorem{rem}[thm]{{\bf{Remark}}}
\numberwithin{equation}{section}
\begin{document}
\newcommand{\cstar}{C^{\ast}}%
%%%%%%%%%%%%%%%%%%%%%%%%%%%
%%%%%%%
\title{On  supersymmetric fermion lattice  systems}
%%%%%%%%
%%%%%%%%%%%%%%%%%%
%%%%%%%%%%%%%%%%%%%%%%%%%%%%%% 
\author{Hajime Moriya}%%
\date{Nov 1 2015}%%%%
\maketitle
\begin{abstract}
This note  provides  a  $\cstar$-algebraic framework  for  supersymmetry. 
Particularly we consider 
  fermion lattice models  satisfying  the simplest    
 supersymmetry relation.  Namely we discuss   a restricted   
 sense of supersymmetry  without a  boson field involved.
 We construct  general supersymmetric 
 $\cstar$-dynamics in terms of a superderivation 
 and a one-parameter group of automorphisms on the CAR-algebra.
(We do not introduce   Grassmann numbers into our formalism.) 
We show  several  basic properties 
 of superderivations  on the fermion lattice system.  
Among others we establish  that superderivations 
  defined on the strictly local algebra  are norm-closable. 
We show  a criterion of superderivations 
 on the fermion lattice system for being  nilpotent.
This criterion can be   easily checked  and hence yields 
 new supersymmetric fermion lattice models.  
\end{abstract}
%%%%%%%
\newpage
%%%%%%
\tableofcontents 
%%%%%
%%%%
\maketitle
%%%%%
\setcounter{tocdepth}{2}
%
%%%
\newcommand{\R}{{\mathbb{R}}}%
\newcommand{\Z}{{\mathbb{Z}}}%
\newcommand{\CC}{{\mathbb{C}}}%
\newcommand{\NN}{{\mathbb{N}}}%
%%%%
\newcommand{\nonum}{\nonumber}%
\newcommand{\unit}{1}%%%
\newcommand{\zeromap}{{\mathbf{0}}}%%%
%%%%
%%%%
\newcommand{\Bl}{{\mathfrak B}}%%%%%%%%%%
\newcommand{\hil}{\mathscr{H}}%
%%%%
\newcommand{\sig}{\sigma}%%%
\newcommand{\vp}{\varphi}%%%%
\newcommand{\lam}{\lambda}%
\newcommand{\ome}{\omega}%%%%
\newcommand{\Lam}{\Lambda}%
\newcommand{\vareps}{\varepsilon}%
%%%%
%%%%
\newcommand{\pot}{{\Phi}}%
\newcommand{\cha}{{\Psi}}%
%%%%
\newcommand{\diam}{{\rm{diam}}}%%%
%%%%
%%%%
\newcommand{\al}{\alpha}%%%%%%
\newcommand{\alt}{\alpha_{t}}%
\newcommand{\als}{\alpha_{s}}%
\newcommand{\alst}{\alpha_{s+t}}%
%%%%
%%%%
\newcommand{\aicr}{a_i^{\,\ast}}%
\newcommand{\ai}{a_i}%
\newcommand{\ajcr}{a_j^{\,\ast}}%
\newcommand{\aj}{a_j}%
\newcommand{\aimcr}{a_{i-1}^{\,\ast}}%
\newcommand{\aim}{a_{i-1}}%
\newcommand{\aipcr}{a_{i+1}^{\,\ast}}%
\newcommand{\aip}{a_{i+1}}%
%%%%
\newcommand{\atwoi}{a_{2i}}%%%%
\newcommand{\atwoicr}{a_{2i}^{\, \ast}}%%%%
\newcommand{\atwoipcr}{a_{2i+1}^{\, \ast}}%%%%
\newcommand{\atwoip}{a_{2i+1}}%%%%%%%%%
\newcommand{\atwoimcr}{a_{2i-1}^{\, \ast}}%%%%
\newcommand{\atwoim}{a_{2i-1}}%%%%%%%%
%%%%
%%%%
\newcommand{\atwoj}{a_{2j}}%%%%
\newcommand{\atwojcr}{a_{2j}^{\, \ast}}%%%%%%
\newcommand{\atwojpcr}{a_{2j+1}^{\, \ast}}%%%%%%
\newcommand{\atwojp}{a_{2j+1}}%%%%%%%%%%%
\newcommand{\atwojmcr}{a_{2j-1}^{\, \ast}}%%%%
\newcommand{\atwojm}{a_{2j-1}}%%%%
%%%%
%%%%
\newcommand{\pio}{\pi_{\ome}}%%%%
\newcommand{\piv}{\pi_{\vp}}%
%%%%
\newcommand{\Gam}{\Gamma}%
\newcommand{\Ome}{\Omega}%
\newcommand{\Omeo}{\Ome_{\ome}}%
\newcommand{\Omev}{\Ome_{\vp}}%
%%%%
%%%%
\newcommand{\Al}{\mathcal{A}}%
\newcommand{\Fl}{\mathcal{F}}%
\newcommand{\core}{\Al_{\circ}}%
%%%%
\newcommand{\Fle}{\Fl_{+}}%
\newcommand{\Flo}{\Fl_{-}}%
\newcommand{\Fleo}{\Fl_{\pm}}%
\newcommand{\Floe}{\Fl_{\mp}}%
%%%%
%%%%
\newcommand{\coree}{{\core}_+}%
\newcommand{\coreo}{{\core}_-}%
\newcommand{\coreeo}{{\core}_\pm}%
\newcommand{\coreoe}{{\core}_\mp}%
%%%%
%%%%
\newcommand{\dcal}{\mathcal{D}}%
\newcommand{\Qast}{Q^{\ast}}%
\newcommand{\Qs}{{\mathcal{Q}}_{{\rm{s}}}}%
\newcommand{\Qsf}{{\mathcal{Q}}_{\rm{s}, 1}}%
\newcommand{\Qss}{{\mathcal{Q}}_{\rm{s}, 2}}%
%%%%
\newcommand{\Qsast}{\Qs^{\ast}}%
\newcommand{\Qsclo}{\overline{\Qs}}%
\newcommand{\Qsfclo}{\overline{\Qsf}}%
\newcommand{\Qssclo}{\overline{\Qss}}%
%%%%
%%%%
\newcommand{\Chasp}{\mathscr{C}}%
\newcommand{\Chaspn}{\mathscr{C}^{\sharp}}%
\newcommand{\Chaspnun}{\mathscr{C}^{\sharp}_{unbroken}}%
%%%%
%%%%
\newcommand{\del}{\delta}%%%%
\newcommand{\deltil}{\tilde{\delta}}%%%%
%%%%
\newcommand{\Derh}{d_0}%%%%
\newcommand{\Derhh}{{d_{0_h}}}%%%%
%%%%
\newcommand{\deln}{\del_n}%%%%
%%%%
\newcommand{\dels}{\del_{\rm{s}}}%%%%
\newcommand{\delsf}{\del_{\rm{s},1}}%%%%
\newcommand{\delss}{\del_{\rm{s}, 2}}%%%%
%%%%
%%%%
\newcommand{\delast}{{\del}^{\ast}}%%%%
\newcommand{\deldoubast}{\del^{\ast \ast}}%%%%
\newcommand{\delsast}{\dels^{\,\ast}}%%%%
%%%%
%%%%
\newcommand{\delcore}{\del|_{\core}}%
\newcommand{\delastcore}{\delast|_{\core}}%
\newcommand{\delscore}{\dels|_{\core}}%
\newcommand{\delsfcore}{\delsf|_{\core}}%
\newcommand{\delsscore}{\delss|_{\core}}%
\newcommand{\Derhcore}{\Derh|_{\core}}%
%%%%
%%%%
\newcommand{\delcl}{\overline{\del}}%
\newcommand{\delastcl}{\overline{\delast}}%
\newcommand{\delscl}{\overline{\dels}}%
\newcommand{\Derhcl}{\overline{\Derh}}%
\newcommand{\delsfcl}{\overline{\delsf}}%
\newcommand{\delsscl}{\overline{\delss}}%
%%%%
%%%%
\newcommand{\delcorecl}{\overline{\delcore}}%
\newcommand{\delscorecl}{\overline{\delscore}}%
%%%%
%%%%
\newcommand{\Derhn}{{\Derh}^{\!n}}%
\newcommand{\Derhk}{{\Derh}^{\!k}}%
\newcommand{\delsn}{{\dels}^{n}}%%%
%%%%
%%%%
\newcommand{\domdel}{\dcal_\del}%
\newcommand{\domdeltil}{\dcal_{\deltil}}%
\newcommand{\domdelast}{\dcal_{\delast}}%
\newcommand{\domdels}{\dcal_{\dels}}%
\newcommand{\domdelsf}{\dcal_{\delsf}}%
\newcommand{\domdelss}{\dcal_{\delss}}%
%%%%
\newcommand{\domdelcl}{\dcal_{\delcl}}%
\newcommand{\domdelastcl}{\dcal_{\delastcl}}%
\newcommand{\domdelsfcl}{\dcal_{\delsfcl}}%
\newcommand{\domdelsscl}{\dcal_{\delsscl}}%
%%%%
\newcommand{\domderh}{\dcal_{\Derh}}%
%%%%
\newcommand{\hilpi}{\hil_{\pi}}%
\newcommand{\hilo}{\hil_{\ome}}%
\newcommand{\hilv}{\hil_{\vp}}%
%%%%
\newcommand{\Znu}{\Z^{\nu}}%
%%%%
\newcommand{\I}{{\mathrm{I}}}%
\newcommand{\J}{{\mathrm{J}}}%
\newcommand{\K}{{\mathrm{K}}}%
\newcommand{\X}{{\mathrm{X}}}%
\newcommand{\Y}{{\mathrm{Y}}}%
\newcommand{\V}{{\mathrm{V}}}%
\newcommand{\C}{{\mathrm{C}}}%
%%%%
\newcommand{\Ic}{\I^{c}}%
\newcommand{\Jc}{\J^{c}}%
\newcommand{\Lamc}{\Lam^{c}}%
%%%%
%%%%
\newcommand{\Cubl}{\C_{l}}%
\newcommand{\Cubr}{\C_{r}}%
\newcommand{\Cubtwor}{\C_{2r}}%
\newcommand{\Cubrone}{\C_{r+1}}%
%%%%
\newcommand{\FlI}{\Fl(\I)}%
\newcommand{\FlX}{\Fl(\X)}%
\newcommand{\FlIc}{\Fl(\Ic)}%
\newcommand{\FlJ}{\Fl(\J)}%
\newcommand{\FlIe}{\Fle(\I)}%
\newcommand{\FlIo}{\Flo(\I)}%
\newcommand{\FlIeo}{\Fleo(\I)}%
\newcommand{\FlIoe}{\Floe(\I)}%
%%%%
\newcommand{\FlXe}{\Fle(\X)}%
\newcommand{\FlXo}{\Flo(\X)}%
\newcommand{\FlJe}{\Fle(\J)}%
\newcommand{\FlJo}{\Flo(\J)}%
%%%%
%%%%
\newcommand{\Ir}{{\hat{\I}}_{r}}%
\newcommand{\Ijr}{{\hat{\I}}_{jr}}%
%%%%%
\newcommand{\Lamr}{{\hat{\Lam}}_{r}}%
\newcommand{\Lamtwor}{{\hat{\Lam}}_{2r}}%
\newcommand{\Itwor}{{\hat{\I}}_{2r}}%
%%%%%
\newcommand{\Itwonr}{{\hat{\I}}_{2nr}}%
\newcommand{\ItwoNr}{{\hat{\I}}_{2Nr}}%
\newcommand{\ItwoNcr}{{\hat{\I}}_{2N_{\circ}r}}%
%%%%%%%%%%%%%%%%%%%%%%%%%%
\newcommand{\Itwokr}{{\hat{\I}}_{2kr}}%
\newcommand{\Itwonminoner}{{\hat{\I}}_{2(n-1)r}}%
\newcommand{\Itwonmintwor}{{\hat{\I}}_{2(n-2)r}}%
%%%%%%%%%%%%%%%%%%%%%%%
\newcommand{\Inr}{{\hat{\I}}_{nr}}%
%%%%%%%%%%%%%%%%%%
\newcommand{\FlIr}{\Fl(\Ir)  }%
\newcommand{\FlIro}{\Flo(\Ir)}%
\newcommand{\FlIre}{\Fle(\Ir)}%
\newcommand{\FlIreo}{\Fleo(\Ir)}%
\newcommand{\FlIroe}{\Floe(\Ir)}%
%%%%
\newcommand{\FlInr}{\Fl(\Inr)  }%
\newcommand{\FlItwonr}{\Fl(\Itwonr)}%
%%%%
%%%%
\newcommand{\FlItwor}{\Fl(\Itwor) }%
\newcommand{\FlItworeo}{\Fleo(\Itwor)  }%
%%%%
\newcommand{\FlIce}{\FlIc^{e}}%
\newcommand{\FlIco}{\FlIc^{o} }%
\newcommand{\FlLam}{\Fl(\Lam) }%
\newcommand{\FlLamz}{\Fl(\Lamz) }%
%%%%
\newcommand{\altlam}{\alt^{\Lam} }%
\newcommand{\alslam}{\als^{\Lam} }%
%%%%
\newcommand{\alchat}{\al^{\!\cha}_t}%
\newcommand{\alchas}{\al^{\!\cha}_s}%
\newcommand{\alchast}{\al^{\!\cha}_{s+t}}%
\newcommand{\alchats}{\al^{\!\cha}_{t+s}}%
\newcommand{\alchanict}{\al^{\!\chanic}_t}%
\newcommand{\alchatrivt}{\al^{\!\chatriv}_t}%
%%%%
%%%%
\newcommand{\altcore}{\alt(\core)}%
\newcommand{\alchatcore}{\alchat(\core)}%
%%%%
\newcommand{\Lamz}{\Lam_{0}}%
\newcommand{\Lamf}{\Lam_{1}}%
\newcommand{\Lams}{\Lam_{2}}%
%%%%
%%%%
\newcommand{\Bp}{B_{+}}%
\newcommand{\Bm}{B_{-}}%
\newcommand{\Bpm}{B_{\pm}}%
\newcommand{\Bast}{B^\ast}%
%%%%
\newcommand{\Bn}{B_{n}}%
\newcommand{\An}{A_{n}}%
\newcommand{\Anz}{A_{{n_0}}}%
\newcommand{\Amm}{A_{m}}%
%%%%
%%%%
\newcommand{\Fp}{F_{+}}%
\newcommand{\Fm}{F_{-}}%
%%%%
\newcommand{\Ap}{A_{+}}%
\newcommand{\Am}{A_{-}}%
%%%%
\newcommand{\Fpm}{F_{\pm}}%
\newcommand{\Gp}{G_{+}}%
\newcommand{\Gm}{G_{-}}%
\newcommand{\Gpm}{G_{\pm}}%
\newcommand{\Gpmd}{{\Gpm}^{\ast}}%
%%%%
%%%%%FUNCTION
\newcommand{\bnlam}{b_{(n,\,\Lam)}}%
\newcommand{\bnlambar}{\bar{b}_{(n,\,\Lam)}}%
\newcommand{\bnlamtil}{\tilde{b}_{(n,\,\Lam)}}%
%%%%
%%%von Neumann algebra%%%%%%%%%%
%%%%
\newcommand{\vn}{{\mathfrak{M}}}%
\newcommand{\vno}{\vn_{\ome}}%
\newcommand{\vnv}{\vn_{\vp}}%
%%%%
%%%%
%%%\left\langle BRAKET
\newcommand{\bra}[1]{\langle #1|}%%%%
\newcommand{\ket}[1]{|#1 \rangle}%%%%			
%%%%
%%%%
\newcommand{\Ran}{{\rm{Ran}}}
\newcommand{\Ad}{{\rm{Ad}}}
%%%%
%%%%
\newcommand{\chaast}{\cha^{\ast}}%
\newcommand{\chasf}{\cha_{\rm{s},1}}%
\newcommand{\chass}{\cha_{\rm{s},2}}%
\newcommand{\chas}{\cha_{\rm{s}}}%
%%%%
%%%%
\newcommand{\chaI}{\cha(\I)}%
\newcommand{\chaJ}{\cha(\J)}%
\newcommand{\chaL}{\cha(\LL)}%
%%%%
\newcommand{\potI}{\pot(\I)}%
%%%%
\newcommand{\delcha}{\del_{\cha}}%
\newcommand{\delchaastnaka}{\del_{\chaast}}%
\newcommand{\delchaast}{{\delcha}^{\!\!\ast}}%
%%%%
\newcommand{\Derhcha}{d_{\cha\, 0}}%%%%
\newcommand{\delchas}{\del_{\chas}}%%%%
\newcommand{\delchasf}{\del_{\chasf}}%%%%
\newcommand{\delchass}{\del_{\chass}}%%%%
%%%%
%%%%
\newcommand{\chanic}{\cha_{\rm{Nic}}}%
\newcommand{\chafen}{\cha_{\rm{Fen}}}%
\newcommand{\chatriv}{\cha_{\rm{trivial}}}%
%%%%
%%%%
\newcommand{\delchanic}{\del_{\chanic}}%
\newcommand{\delchafen}{\del_{\chafen}}%
\newcommand{\delchatriv}{\del_{\chatriv}}%
\newcommand{\delchatrivast}{{\delchatriv}^{\!\!\ast}}%
%%%%
%%%%
\newcommand{\Derhchanic}{d_{\chanic\, 0}}%%%%
\newcommand{\Derhchatriv}{d_{\chatriv\, 0}}%%%%
%%%%
%%%%
\newcommand{\chanicI}{\chanic(\I)}%
\newcommand{\chanicJ}{\chanic(\J)}%
%%%%
%%%%
\newcommand{\delchacl}{\overline{\delcha}}%
\newcommand{\delchaastcl}{\overline{\delchaast}}%
\newcommand{\delchascl}{\overline{\delchas}}%
\newcommand{\delchasfcl}{\overline{\delchasf}}%
\newcommand{\delchasscl}{\overline{\delchass}}%
%%%%
%%%%
\newcommand{\domdelcha}{\dcal_{\delcha}}%
\newcommand{\domdelchaast}{\dcal_{\delchaast}}%
\newcommand{\domdelchas}{\dcal_{\delchas}}%
\newcommand{\domdelchasf}{\dcal_{\delchasf}}%
\newcommand{\domdelchass}{\dcal_{\delchass}}%
%%%%
%%%%
\newcommand{\domdelchacl}{\dcal_{\delchacl}}%
\newcommand{\domdelchaastcl}{\dcal_{\delchaastcl}}%
\newcommand{\domdelchascl}{\dcal_{\delchascl}}%
\newcommand{\domdelchasfcl}{\dcal_{\delchasfcl}}%
\newcommand{\domdelchasscl}{\dcal_{\delchasscl}}%
%%%%
%%%%
\newcommand{\delchaslam}{\del_{\chas,\,\Lam}}%%%%
\newcommand{\Derhchalam}{d_{\cha\,0,\, \Lam}}%
\newcommand{\Derhchalamn}{{d_{\cha\,0,\, \Lam}}^{\!\!\!n}}%
\newcommand{\Derhchalamk}{{d_{\cha\,0, \,\Lam}}^{\!\!\!k}}%
%%%%
%%%%
\newcommand{\Derhchan}{{\Derhcha}^{\!\!n}}%
\newcommand{\Derhchanmone}{{\Derhcha}^{\!\!n-1}}%
%%%
\newcommand{\Derhchak}{{\Derhcha}^{\!\!k}}%
\newcommand{\delchasn}{{\delchas}^{\!\!\!\!n}}%%%
%%%%
\newcommand{\isan}{\{ 2i-1,\, 2i,\, 2i+1 \}}%%%%%
\newcommand{\iryodonari}{\{ i-1,\, i,\, i+1 \}}%%%%%
%%%%
%%%%
\newcommand{\Flisan}{\Fl\bigl(\isan\bigr)}%%%%
\newcommand{\Flisano}{\Flo\bigl(\isan\bigr)}%%%%
\newcommand{\chaisan}{\cha\bigl(\isan\bigr)}%%%%%
\newcommand{\chanicisan}{\chanic\bigl(\isan\bigr)}%%%%%
\newcommand{\chafeni}{\chafen\bigl(\iryodonari\bigr)}%%%%%
\newcommand{\chafenI}{\chafen\bigl(\I\bigr)}%%%%%
\newcommand{\chafenisingle}{\chafen\bigl(\{ i \}\bigr)}%%%%%
%%%%
\newcommand{\chatrivi}{\chatriv\bigl(\{ i \}\bigr)}%%%%%
\newcommand{\chatrivasti}{\chatriv^{\ast}\bigl(\{ i \}\bigr)}%%%%%
%%%%
\newcommand{\chatrivI}{\chatriv\bigl(\I\bigr)}%%%%%
%%%%
%%%%
\newcommand{\Lchas}{l_{\chas}}%
\newcommand{\Lchasn}{{\Lchas}^{\!\!n}}%
\newcommand{\kchas}{k_{\left\{A,\, \chas\right\}}}%%%%
\newcommand{\kpchas}{k^{\prime}_{\left\{A,\, \chas\right\}}}%%%%
\newcommand{\Mchas}{m_{\chas}}%%%%
\newcommand{\Mchasn}{{\Mchas}^{\!\!n}}%
%%%%
\newcommand{\chax}{\cha(\X)}%
\newcommand{\chaxone}{\cha(\X_1)}%
\newcommand{\chaxtwo}{\cha(\X_2)}%
\newcommand{\chaxn}{\cha(\X_n)}%
%%%%
\newcommand{\chaastI}{\chaast(\I)}%
\newcommand{\chaastJ}{\chaast(\J)}%
%%%%
\newcommand{\Vzero}{\V_{0}}%
\newcommand{\Vone}{\V_{1}}%
\newcommand{\Vi}{\V_{i}}%
\newcommand{\Vj}{\V_{j}}%
\newcommand{\Vjmone}{\V_{j-1}}%
\newcommand{\Vn}{\V_{n}}%
\newcommand{\Vnmone}{\V_{n-1}}%
%%%%
\newcommand{\mugen}{\Lam \nearrow \Znu}%%
\newcommand{\Lamlim}{\lim_{\mugen}}%%
\newcommand{\nlim}{\lim_{n\to\infty}}%%
%%%%
\newcommand{\delk}{\del_k}%%%%
\newcommand{\omen}{\ome_n}%%%%
\newcommand{\omek}{\ome_k}%%%%
%%%%
%%%%

%%%%
\section{Introduction}
\label{sec:INTRO}
It has been  expected 
 that supersymmetry will play  a crucial role      
 in the  unification of fundamental interactions in  particle physics,
 see e.g. \cite{WEIN}.
Furthermore,   the concept of supersymmetry  has 
 been influential and fruitful
 in  wide range of areas in physics and mathematics. 
Among many  topics related to supersymmetry,  we shall refer to 
 supersymmetric quantum mechanics (SUSY QM)   
  in which we find  remarkable  
 interplays between mathematics and physics.
  For general account of  SUSY QM, let us refer to
 a comprehensive work  \cite{COOP} and the reference therein.  
In this note we  study  fermion lattice models with   hidden 
 supersymmetry. A  spinless
 fermion lattice model that has  a hidden  supersymmetry 
 was   proposed  
by Nicolai \cite{NIC} in 1976;
 this  quantum statistical-mechanical model 
is another (unfortunately not well-known)  example of 
supersymmetric quantum mechanics.{\footnote{The SUSY QM model  
 introduced by  Witten  \cite{WITT81, WITT82} is  a classic 
in supersymmetry theory. 
This   model  by Nicolai 
 is  another pioneer work of  
 non-relativistic SUSY, see  \cite{JUN}.}}
Recently, other supersymmetric fermion lattice 
 models have been proposed and investigated   
by  Fendly et al. \cite{FEN1}. 

There have  been extensive works  of $\cstar$-algebraic 
 quantum field theory and quantum statistical mechanics, 
 for general references we refer to 
\cite{BR} \cite{HAAG}.
 Although 
it is  straightforward  to write down  
a  supersymmetry algebra  (without central charges)  
on  a graded $\cstar$-algebra
 {\it{heuristically}}  as found  in  some previous  works 
\cite{JAFLOK, JAFdeform, KAST},
 it is  not clear  whether and how 
supersymmetric models  can be formulated in $\cstar$-algebraic 
 quantum theory  or  its natural extension. 
It seems that   construction of  supersymmetry dynamics 
  within a graded  $\cstar$-algebra  is unexpectedly difficult 
 unless the algebra is finite dimensional. 

To make  steps toward   supersymmetry theory in  $\cstar$-algebras,
 we shall focus on  fermion lattice systems.
 Based on   a  $\cstar$-algebraic 
 framework  for  supersymmetric fermion lattice models
 that we will show 
 we  rigorously  discuss  supersymmetric dynamics
 and supersymmetric states in the infinite volume limit.
We have to emphasize  that supersymmetry originally means  a symmetry
  between fermions and bosons.
  Buchholz-Grundling have 
 invented a  $\cstar$-algebra approach  
 to  supersymmetry between  fermions and bosons
  in   \cite{BUGR}. 
 On the other hand, this note only  deals with 
fermions. 
By  exploiting such a simplified situation with no  boson field 
 we give  a general class of  $\cstar$-dynamics  on the CAR algebra.
 This  corresponds to  supersymmetric  fermion lattice models 
 of finite-range interactions.

Let us  explain the plan  of this paper.
In  Sect.\ref{sec:NOTATION}
 we  introduce   superderivations 
 of  a  graded $\cstar$-algebra. 
We define supersymmetric states
 and show their basic  properties.
In Sect.\ref{sec:overview} we provide a heuristic overview  
 on supersymmetric fermion  lattice systems.
In Sect.\ref{sec:SUSYfer} we formulate 
supersymmetric fermion models of finite-range interactions
 as strongly continuous $\cstar$-dynamics on the CAR algebra.    
A superderivation on  the fermion lattice system 
 is determined  by assigning local fermionic charges over the lattice. 
This  is analogous  to the  well-known construction 
 of a  time generator generated by  local Hamiltonians 
over the lattice, see \cite{BR}.
 We will show that these 
 superderivations on the CAR algebra are  norm-closable 
when  the associated  supersymmetry is unbroken.
(It is not known, however,  
 whether this statement holds for the case of  
broken  supersymmetry. We conjecture that this is still correct 
 when the supersymmetry is broken.) 
Using  the norm-closability of superderivations
 mentioned above 
 we  establish a rigorous formulation  of supersymmetry  
that includes  commutativity   
between  those  superderivations and  the global time evolution
 generated by them.
In Sect.\ref{sec:EX}
we give  some concrete supersymmetric fermion lattice  models  
 in our  $\cstar$-algebraic framework of   Sect.\ref{sec:SUSYfer}.
These are based on  the  model by  Nicolai \cite{NIC} and 
 the model by  Fendly et al. \cite{FEN1}.
Sect.\ref{sec:ABS} is  a summary of this note. 
We present a set of  axioms  for supersymmetric 
$\cstar$-dynamical systems which  is  abstracted from the supersymmetric  
$\cstar$-dynamics on the CAR algebra of  Sect.\ref{sec:SUSYfer}.

\section{Notation}
\label{sec:NOTATION}
\subsection{Superderivations} 
\label{subsec:superderivation}
We first  introduce 
 superderivations in a  general $\cstar$-algebra.
 Let  $\Fl$ denote  a graded $\cstar$-algebra  with a 
grading  $\gamma$, where  $\gamma$ is given by a $\Z_{2}$-group of 
$\ast$-automorphisms  of $\Fl$.  The graded structure 
of $\Fl$ induced by $\gamma$  is as follows:    
\begin{equation}
\label{eq:EO}
\Fl=\Fle\oplus\Flo,\ \ 
 \Fle:= \{F \in \Fl \; \bigl| \;   \gamma(F)=F  \},\quad  
 \Flo:= \{F \in \Fl  \; \bigl| \;  \gamma(F)=-F  \}. 
 \end{equation}
The above   $\Fle$ and $\Flo$ are called  
the even and odd parts of $\Fl$,  respectively.  
The  graded commutator 
$[\; \ , \; \ ]_{\gamma}$  is defined
  on $\Fl$ as
\begin{align}
\label{eq:gcom}
[\Fp, \;  G]_{\gamma} &= [\Fp, \;  G]
=  \Fp G - G \Fp  {\text {\ \ for \ }}
\Fp \in \Fle,    \ G \in \Fl, \notag\\
[F, \;  \Gp]_{\gamma} &= [F, \;  \Gp]
 =F \Gp - \Gp  F {\text {\ \ for \ }}
F \in \Fl, \ \Gp \in \Fle, \notag\\
[\Fm, \;  \Gm]_{\gamma} &= \{ \Fm  \; \Gm \}=
\Fm  \Gm +\Gm \Fm   {\text {\ \ for \ }}
\Fm \in \Flo, \ \Gm \in \Flo. 
\end{align}

Let  $\core$  be   a  {\it{globally}} $\gamma$-invariant  $\ast$-subalgebra 
 of $\Fl$. Namely  
each element of $\core$ is not necessarily $\gamma$-invariant,
 however, $\gamma(\core)=\core$ holds. 
A linear map  $\del: \core  \mapsto \Fl$ 
  is called  a {\it{superderivation}} of  $\Fl$  with respect to  $\gamma$ if 
 it  is  {\it{odd}}  with respect to the grading: 
\begin{equation}
\label{eq:oddsuperder}
\del\cdot \gamma=-\gamma \cdot \del\ \ 
{\text {on}}\ \core, 
\end{equation}
and  it satisfies  the {\it{graded Leibniz rule}}:
\begin{equation}
\label{eq:gleib}
\del(AB)=
\del(A)B+\gamma (A) \del(B)\ \  {\text {for every}}\ A, B\in \core.
\end{equation}
By \eqref{eq:oddsuperder}
\begin{equation}
\label{eq:oddmap}
\del(\coree)\subset\Flo,\ \ 
\del(\coreo)\subset\Fle,
\end{equation}
where 
\begin{equation}
\label{eq:coreevenodd}
\coree:=\core \cap\Fle,\ \ 
 \coreo:=\core \cap\Flo.
\end{equation}
Assume that  
  $\Fl$  has  a  unit  element  $\unit\in \Fl$
 and the subalgebra $\core$ includes this unit.
Since  $\del(\unit)=\del(\unit\cdot\unit)=\del(\unit)\unit
+\unit\del(\unit)=2\del(\unit)$,
we have   
\begin{equation}
\label{eq:delunitzero}
\del(\unit)=0.
\end{equation}

For  a superderivation $\del$ defined on $\core$
  its  conjugation  superderivation is defined  by 
\begin{equation}
\label{eq:delastDEF}
\delast(A):=- \left(\del\bigl(\gamma (A^{\ast}) \bigr)\right)^{\ast}
\quad \text{for every}\ \ A  \in \core,
\end{equation}
where  $^{\ast}$ on  the right hand side 
denotes  the ${\ast}$-operation  of  the $\cstar$-algebra $\Fl$.
It is easy  to see that  $\delast:\core  \mapsto \Fl$ 
is a superderivation  and that 
\begin{equation}
\label{eq:astast}
\deldoubast=\del. 
\end{equation}
If a  superderivation $\dels$ 
 is `symmetric' with respect to the $\ast$-operation
 on superderivations  defined in \eqref{eq:delastDEF}, 
\begin{equation}
\label{eq:hermitedelDEF}
 \dels=\delsast \ \ \text{on}\  \core,
\end{equation}
 then it is said to be  {\it{hermite}}.

For a general superderivation $\del$ of $\Fl$ defined on 
 $\core$, we introduce  the following pair of   hermite  superderivations:
\begin{equation}
\label{eq:dels-futatu}
 \delsf:=\del+\delast, \quad  
\delss:=i(\del-\delast) \ \ \text{on}\  \core.
\end{equation}
Conversely, 
\begin{equation}
\label{eq:del-gyaku}
\del=\frac{1}{2}(\delsf-i\delss), \quad
 \delast=\frac{1}{2}(\delsf+i\delss) \ \text{on}\  \core.
\end{equation}

\begin{rem}
\label{rem:Grassmann}
 We do not introduce Grassmann numbers (infinitesimal fermionic 
 c-number spinors) into the  $\cstar$-algebra
 unlike the  superfield formalism  which is  a basic language 
for  supersymmetric field theories \cite{WEIN}.
\end{rem}

\begin{rem}
\label{rem:KIGOUdeldels}
We denote a  general superderivation by $\del$, and a general 
hermite superderivation by $\dels$. Hence  $\dels$  stands for 
 both $\delsf$ and $\delss$. 
Later $\del$ is assumed to be  nilpotent (accordingly 
 to be non-hermite).
 \end{rem}

\subsection{Supersymmetric  states}
\label{subsec:susystates}
We specify the meaning of  `supersymmetric states'.
\begin{defn}
\label{defn:SUSYSTATE}
Let $\del$ be a superderivation of $\Fl$ and let  $\core$ denote  
its domain subalgebra (which is  
a unital globally  $\gamma$-invariant  
$\ast$-subalgebra of $\Fl$ by definition).   
If a  state  $\vp$ on $\Fl$  is  invariant  under  $\del$, 
namely 
\begin{equation}
\label{eq:susyinv}
\vp \left(\del(A)\right)=0\ \ {\text{for every}}\  A\in \core,
\end{equation}
then it is said to be  supersymmetric (with respect to $\del$).
\end{defn}

In the above definition, we 
  do not need  an   actual supersymmetry relation 
 that involves  time evolution.
It requires only a superderivation. 
The following two statements are  obvious.
\begin{prop}
\label{prop:susy-con}
If a state $\vp$ on  $\Fl$ 
is  supersymmetric with respect to a superderivation 
$\del$ defined on $\core$, 
then it is also  supersymmetric with respect to $\delast${\rm{:}}
 \begin{equation}
\label{eq:susyinvast}
\vp \left(\delast(A)\right)=0\ {\text{for every}}\  A\in \core.
\end{equation}
\end{prop}
\begin{proof}
By  \eqref{eq:delastDEF} and the $\gamma$-invariance of $\core$,  
Eq.\eqref{eq:susyinv} implies Eq.\eqref{eq:susyinvast} (and vice versa).
\end{proof}

\begin{prop}
\label{prop:inv-symdel}
Let $\del$ be  a superderivation defined on a globally 
 $\gamma$-invariant  
$\ast$-subalgebra $\core$ of   $\Fl$.
A state  $\vp$ on $\Fl$ is supersymmetric with respect to $\del$  
if and only if it is invariant under  
  each of  $\delsf$ and  $\delss$, where $\delsf$ and  $\delss$ 
denote  the hermite superderivations  on $\core$ given in   
\eqref{eq:dels-futatu}.
\end{prop}

\begin{proof}
By  \eqref{eq:dels-futatu} \eqref{eq:del-gyaku} 
 Proposition \eqref{prop:susy-con} implies  the assertion.
\end{proof}

We then  specify  `unbroken--broken supersymmetry'. 
\begin{defn}
\label{defn:unbroken--broken}
Let $\del$ be a  superderivation of  $\Fl$.
If there exists  a supersymmetric  state   on $\Fl$
 with respect to $\del$ as in  Definition \eqref{defn:SUSYSTATE},
 then it is said that the supersymmetry  (generated  by  $\del$) is unbroken.
If no such state exists,  then it is said that the supersymmetry is 
 spontaneously  broken.
\end{defn}

\begin{rem} 
\label{rem:unbroken--broken}
 In  Definition \eqref{defn:unbroken--broken}
 we do not  require   time evolution  satisfying 
 the supersymmetry relation.
 Hence  Definition \eqref{defn:unbroken--broken}
  is valid not only for  the  (usual) dynamical supersymmetry
 but also for the kinematical supersymmetry which refers to  
 a more general  fermion  symmetry. 
In this note   we will deal with  
 only the dynamical supersymmetry.
\end{rem}

Later the following topological  property of superderivations
 is important. 
\begin{defn}
\label{defn:normclosable}
A  superderivation $\del$ 
defined on a norm dense domain $\core$ 
 of $\Fl$ is called norm-closable if 
for  any sequence  $\{\An \in \core\}$, the  
 convergence  $\lim_{n\to \infty} \An = 0$ and  
$\lim_{n\to \infty} \del(\An) =B$ in norm implies that  $B=0$.
If $\del$ is  norm-closable, then 
its least closed extension  
  $\delcl$ is called the closure of $\del$.
\end{defn}

\begin{prop}
\label{prop:ast-normclo}
If a superderivation $\del$ of $\Fl$ defined on 
a norm-dense globally $\gamma$-invariant  
$\ast$-subalgebra $\core$
 is  norm-closable, then its conjugate  
superderivation $\delast$  on $\core$ is also norm-closable.
\end{prop} 
\begin{proof}
Note that for any $\{\An \in \core\}$ 
 convergence $\lim_{n\to \infty} \An = 0$  and
convergence  $\lim_{n\to \infty} \gamma(\An^{\ast}) = 0$ are equivalent.
 Hence by   the  form  of $\delast$ as given in  \eqref{eq:delastDEF}
 the statement  is satisfied.
\end{proof}

 The following  statement derives   norm-closability 
 of superderivations when the associated 
 supersymmetry is unbroken  under a  general $\cstar$-algebraic 
setting. 
It is obviously applicable to fermion lattice systems.
\begin{prop}
\label{prop:faithclo}
Let $\Fl$ be a unital graded $\cstar$-algebra 
and  let $\gamma$ denote its grading automorphism. 
Let $\del$ be a superderivation defined on  
 a globally $\gamma$-invariant  $\ast$-subalgebra $\core$ of $\Fl$.
Suppose that  $\core$ is norm-dense in $\Fl$.
Suppose that there exists a supersymmetric state $\vp$  on $\Fl$ 
with respect to $\del$ as in  Definition 
{\rm{\eqref{defn:SUSYSTATE}}} and that 
its GNS representation $\bigl(\piv,\; \hilv,\; \Omev \bigr)$  
gives a faithful representation of $\Fl$.
  Then $\del$ is norm-closable.
\end{prop}

\begin{proof}
As in   \cite{BU} 
let us introduce   
\begin{equation}
\label{eq:Qdef}
Q \piv(A)  \Omev:= \piv \bigl(\del(A)\bigr)\Omev,  \ \; A\in \core.
\end{equation}
This gives  a well-defined   
  closable linear operator $Q$ on $\hilv$,
 as its adjoint $\Qast$ is  defined also on the
 norm-dense subspace $\piv(\core)\Omev$ by   
\begin{equation*}
\Qast \piv(A)  \Omev:= \piv \bigl(\delast(A)\bigr)\Omev,  \ \; A\in \core.
\end{equation*}
It is easy to see that the following 
operator equality
holds  for every $A\in \core$
\begin{equation}
\label{eq:Qoponcore}
\piv \bigl(\del(A)\bigr) = Q \piv(A)- 
\piv\bigl(\gamma(A)\bigr)Q
  \ \ {\text {on}}\ \piv(\core)\Omev. 
\end{equation}
Let  $\{\An\in \core  \}_{n\in\NN}$ be a sequence such that 
\begin{equation*}
\An  \to 0 \ \  
{\text{and}}\  \ 
\del(\An ) \to B \in \Fl 
\ \  {\text{in norm as}}\ n\to\infty. 
\end{equation*}
This obviously yields  the convergence   
 $\gamma(\An)\to 0$ in norm.
In the GNS representation 
 $\bigl(\piv,\; \hilv,\; \Omev  \bigr)$ for any supersymmetric 
 state $\vp$,  
we have for every  $C, D \in \core$
\begin{align*}
& \left( \piv(D)\Omev,\  \piv(B) \piv(C) \Omev\right) \nonum\\
 &=\lim_{n\to\infty}
\left( \piv(D)\Omev,\   \piv\bigl(\del(\An)  \bigr)   \piv(C)  \Omev\right)
\nonum \\
 &=\lim_{n\to\infty}
\left( \piv(D)\Omev,\  Q \piv (\An  )   \piv(C)  \Omev\right)
-\lim_{n\to\infty}
\left( \piv(D)\Omev,\  \piv( \gamma(\An)  ) Q  \piv(C)  \Omev\right)\nonum\\
\nonum \\
 &=\lim_{n\to\infty}
\left(\Qast \piv(D)\Omev,\   \piv (\An  )   \piv(C)\Omev\right)
-\lim_{n\to\infty}\left( \piv(D)\Omev,\  \piv( \gamma(\An)  ) Q \piv(C) \Omev\right)\nonum\\
&=0-0=0,
\end{align*}
where the operator identity  \eqref{eq:Qoponcore},
 and then the norm convergence   $\piv( \An  )\to 0$ and 
$\piv( \gamma(\An)  )\to 0$ in norm is  noted.
As $\piv(\core)\Omev$  is dense in $\hilv$, this yields $\piv(B)=0$.
 Since   $\piv$ is injective by the assumption, we conclude that  $B=0$. 
 The proof  is  completed.
\end{proof}

We are interested in  the status of 
unbroken-broken  supersymmetry in the infinite volume limit.
Let us  quote  the following well-known   statement  from 
 \cite{WITT82}. \\

\noindent{\it{If supersymmetry is unbroken
 in an arbitrary finite volume $V$, this means that 
 the ground-state energy $E(V)$ is zero for every $V$.  
Since the large-$V$ limit of zero is zero,
 this means that the ground-state energy is zero in the infinite-volume 
 limit, and that supersymmetry is unbroken in this limit.}}\\

 The following proposition provides 
  a  rigorous  derivation  of the above statement.
By this one  can  show the existence of  supersymmetric states 
in the infinite-volume limit  for  some concrete models.

\begin{prop}
\label{prop:infinite-volume}
Let $\del$ be a  superderivation on a graded $\cstar$-algebra $\Fl$
 and let $\core$ denote its domain.
Assume that there exists a sequence of superderivations 
$\{\deln\}$ on the same domain $\core$ 
such  that 
 \begin{eqnarray}
\label{eq:appdel}
\del(A)=\lim_{n}\deln(A)
{\mbox {\ \ in norm for each}}\ A\in \core.
\end{eqnarray}
Assume that the supersymmetry 
generated by  each $\deln$
is unbroken, namely for  each  $\deln$
there exits  an invariant  state.
Then  the supersymmetry 
 generated by  $\del$ is unbroken.
\end{prop}
\begin{proof}
We will  provide  a state 
  on $\Fl$ which is invariant under the superderivation  $\del$.
Let  $\omen$ denote  a state on $\Fl$ which is invariant under $\deln$.
As the state space of a $\cstar$-algebra is compact in the 
weak-$\ast$ topology, there exists a state which is a cluster 
point of the sequence $\{\omen\}$ in the weak-$\ast$ topology.
Let $\ome$ denote any of such cluster points of $\{\omen\}$.
We will show that $\ome$  gives   a desired state. 

Take an arbitrary $A\in\core$.
We fix  an arbitrary  $\varepsilon>0$.
By the assumption \eqref{eq:appdel} 
there exists $k_0$ such that 
for every $k\ge k_0$
\begin{align*}
\left\|\del(A)-\delk(A) \right\|<\varepsilon/2. 
\end{align*}
As  $\ome$ is a cluster point of $\{\omen\}$
 in the weak-$\ast$ topology, we can take a subsequence 
  $\{n^{\prime}\}$ of $\{n\}$ such that 
\begin{eqnarray*}
\ome \left(\del(A)\right)
=\lim_{n^{\prime}}\ome_{n^{\prime}} \left(\del(A)\right).
\end{eqnarray*}
Hence by passing to 
the subsequence $\{n^{\prime}\}$  from the original 
sequence $\{n\}$ 
there exists $k_1$ such that 
for every $k\ge k_1$
\begin{align*}
\left|\ome \left(\del(A)\right)-\omek \left(\del(A)\right) \right|
<\varepsilon/2. 
\end{align*}
For any $n$ (in the chosen sequence
 $\{n^{\prime}\}$) such that  $n\ge \max{\{k_0, k_1\}}$, we have  
\begin{align*}
&\Bigl|\ome \left(\del(A)\right)-\omen \left(\deln(A)\right) \Bigr|
=\Bigl|\ome \left(\del(A)\right)
-\omen \left(\del(A)\right)
+\omen \left(\del(A)\right)
-\omen \left(\deln(A)\right) \Bigr|
\nonum\\
&\le \left|\ome \left(\del(A)\right)
-\omen \left(\del(A)\right)
\right|+\left|\omen\left( \del(A)-\deln(A)\right)\right|\nonum\\
&\le \left|\ome \left(\del(A)\right)
-\omen \left(\del(A)\right)
\right|+\left\| \del(A)-\deln(A) \right\|
<\varepsilon/2+\varepsilon/2=\varepsilon. 
\end{align*}
Since  $\omen \left(\deln(A)\right)=0$ by the assumption, we have 
$\left|\ome \left(\del(A)\right) \right|<\varepsilon$.
As  $\varepsilon>0$ is arbitrary, we have 
$\ome \left(\del(A)\right)=0$.
Therefore $\ome$ is  supersymmetric  
 with respect to $\del$. 
\end{proof}

\begin{rem} 
\label{rem:infinite-volume}
 The convergence condition 
  \eqref{eq:appdel} 
 in Proposition \eqref{prop:infinite-volume}
 may  denote  the infinite-volume limit of finite subsystems.
Proposition \eqref{prop:infinite-volume} can be applied to 
fermion  lattice systems and quantum spin lattice systems.
\end{rem}

\section{Overview}
\label{sec:overview}
 We shall provide  a heuristic overview  on 
 supersymmetric  fermion lattice models.
For simplicity let us 
consider  the  one-dimensional lattice  $\Z$.
Let $\ai$ and $\aicr$ denote the annihilation operator 
and the creation operator of a spinless fermion at a site $i\in \Z$. 
 Those satisfy  the canonical anticommutation relations:
\begin{align*}
\{ \aicr, \aj \}&=\delta_{i,j}\, \unit, \nonumber \\
\{ \aicr, \ajcr \}&=\{ \ai, \aj \}=0.
\end{align*}
In \cite[Sect.3]{NIC}{\footnote{In \cite[Sect.3]{NIC}  
the half-sided lattice $\NN$ 
is considered  instead of the lattice $\Z$.}}
  the following supercharge  
\begin{align}
\label{eq:Q-intro}
Q&:=\sum_{i\in \Z}  a_{2i+1} a^{\ast}_{2i} a_{2i-1},
\end{align}
and  the Hamiltonian 
\begin{align}
\label{eq:H-intro}
H&:=\{ Q,\; {Q}^{\ast} \}\nonum\\
&=\sum_{i\in \Z}\bigl\{ 
a^{\ast}_{2i}a_{2i-1}a_{2i+2}a^{\ast}_{2i+3}
+a^{\ast}_{2i-1}a_{2i}a_{2i+3}a^{\ast}_{2i+2} \nonum\\
&\ +a^{\ast}_{2i}a_{2i}a_{2i+1}a^{\ast}_{2i+1}
+a^{\ast}_{2i-1}a_{2i-1}a_{2i}a^{\ast}_{2i}
- a^{\ast}_{2i-1} a_{2i-1}a_{2i+1}a^{\ast}_{2i+1} \bigr\}
\end{align}
are introduced.
Let $N$ denote the total  fermion number operator:
\begin{equation*}
N:=\sum_{i\in \Z}  a^{\ast}_{i} a_{i}.
\end{equation*}
 We see  that 
\begin{equation}
\label{eq:Qodd-intro}
\{(-1)^{N},\ Q\}= \{(-1)^{N},\ Q^{\ast}\}=0.
\end{equation}
Namely  $Q$ and $\Qast$ are  fermionic.
It is straightforward  to  see that
 the supercharge  $Q$  is  nilpotent:     
\begin{align}
\label{eq:Qnil-intro}
Q^{2}=0.
\end{align}
From  \eqref{eq:H-intro} and  \eqref{eq:Qnil-intro}
 the  commutativity  between the Hamiltonian and the 
 supercharges  follows:
\begin{equation}
\label{eq:HQkakan}
[H,\ Q]=[H,\ Q^{\ast}]=0.
\end{equation}  
The pair of self-adjoint (real) supercharges are given by{\footnote{The subscript `{\rm{s}}' of  
$\Qsf$ and $\Qss$ will denote   `self-adjoint'  or `symmetric' 
operators on a Hilbert space.}}
\begin{equation}
\label{eq:Qsfdef}
\Qsf:=Q+Q^{\ast},\ \ \Qss:=i(Q-Q^{\ast}).
\end{equation}  
Inverting  the above relations we have   
\begin{equation}
\label{eq:Qdecomp}
Q=\frac{1}{2}\left(\Qsf-i\Qss\right),\ \ 
Q^{\ast}=\frac{1}{2}\left(\Qsf+i\Qss\right). 
\end{equation}  
By definition
\begin{equation}
\label{eq:Qsfsym}
{\Qsf}^{\!\!\ast}=\Qsf,\ \ 
{\Qss}^{\!\!\ast}=\Qss,  
\end{equation}  
\begin{equation}
\label{eq:Qsfodd}
\{(-1)^{N},\ \Qsf\}= \{(-1)^{N},\ \Qss\}=0,
\end{equation}
\begin{equation}
\label{eq:Qs-anticomm}
\{\Qsf,\ \Qss\}=0, 
\end{equation}
\begin{equation}
\label{eq:susysymH}
H=\Qsf^{\;2}=\Qss^{\;2}, 
\end{equation}
and 
  \begin{equation}
\label{eq:HQskakan}
[H,\ \Qsf]=[H,\ \Qss]=0.
\end{equation}
The above set of relations 
in terms of $\Qsf$, $\Qss$ and  $H$
gives an  equivalent  expression of the supersymmetry algebra. 

Note that   the  supercharges  $Q$, $Q^{\ast}$, 
 $\Qsf$, $\Qss$ and   the Hamiltonian $H$
 given above  are not well defined as  linear operators.
However, they determine well-defined  infinitesimal generators. 
Let $\core$  denote  the  local  algebra 
which is  generated by all  local elements.
Then let us define the superderivation 
\begin{equation}
\label{eq:del-intro}
\del(A):=[Q,\; A]_{\Gamma}
{\text {\ \ for every  \ }} A \in \core, 
\end{equation}
where the symbol $[\ , \ ]_{\Gamma}$ denotes the  
graded commutator as
in \eqref{eq:gcom} with the grading 
 automorphism $\Gamma:=\Ad{(-1)^{N}}$.
We note that 
\begin{equation}
\label{eq:delast-intro}
\delast(A):=[Q^{\ast},\; A]_{\Gamma}
{\text {\ \ for every  \ }} A \in \core. 
\end{equation}
By using the (usual) commutator  we define
 the derivation which gives an infinitesimal 
time-generator  
\begin{equation}
\label{eq:Derh-intro}
\Derh(A):=[H,\; A]
{\text {\ \ for every  \ }} A \in \core. 
\end{equation}
From  \eqref{eq:Q-intro} \eqref{eq:H-intro} \eqref{eq:Qnil-intro} 
 \eqref{eq:del-intro} \eqref{eq:delast-intro} \eqref{eq:Derh-intro}
we obtain 
the  supersymmetry algebra  
in terms  of the superderivations 
 and the time-derivation:
\begin{equation}
\label{eq:delnil-intro}
\del\cdot\del=\zeromap
{\text {\ \ on\ \ }}   \core,
\end{equation} 
\begin{equation}
\label{eq:susy-intro}
\Derh=\delast\cdot\del+\del\cdot\delast  
{\text {\ on   \ }}  \core.
\end{equation}
Here we   note that 
\begin{equation}
\label{eq:DALGsoboku}
\del(\core)\subset \core,\ 
 \delast(\core)\subset \core,\ 
\Derh(\core)\subset \core.
\end{equation}

Let  
\begin{align}
\label{eq:dels-intro}
\delsf(A)&:=[\Qsf,\; A]_{\Gamma}
{\text {\ \ for every  \ }} A \in \core,\nonum\\ 
\delss(A)&:=[\Qss,\; A]_{\Gamma}
{\text {\ \ for every  \ }} A \in \core.
\end{align}
Then  the supersymmetry algebra 
expressed by Eqs.\eqref{eq:delnil-intro} 
\eqref{eq:susy-intro}
is rewritten as 
\begin{equation}
\label{eq:symdel-intro}
\delsf\cdot\delss+\delss\cdot \delsf=\zeromap
{\text {\ \ on\ \ }}
\core,  
\end{equation}
\begin{equation}
\label{eq:susy-sym-intro}
\Derh={\delsf}^{\!\!\!2}
={\delss}^{\!\!\!2} {\text {\ \ on \ \ }} \core.
\end{equation}
Here we  note that 
\begin{equation}
\label{eq:DALGsobokuII}
\delsf(\core)\subset \core,\ 
 \delss(\core)\subset \core,\ 
\Derh(\core)\subset \core.
\end{equation}

From  \eqref{eq:HQkakan} 
 \eqref{eq:del-intro} \eqref{eq:delast-intro} and \eqref{eq:Derh-intro}
the following commutativity relations hold:
\begin{equation}
\label{eq:koukan-ferntro}
\del\cdot \Derh=\Derh\cdot \del,\ \ 
\delast\cdot \Derh=\Derh\cdot \delast\ 
 {\text{\ on\ \ }}\core.
\end{equation}
Similarly from 
 \eqref{eq:HQskakan} 
 \eqref{eq:dels-intro} and  \eqref{eq:Derh-intro}
the following commutativity relations hold:
\begin{equation}
\delsf\cdot \Derh=\Derh\cdot \delsf,\ \ 
\delss\cdot \Derh=\Derh\cdot \delss\ 
 {\text{\ on\ \ }}\core.
\end{equation}

\begin{rem} 
\label{rem:Q-broken-unbroken}
If a global supersymmetry  is  spontaneously broken,
 then the corresponding supercharge  does not exist  as 
 a densely defined linear operator,  
cf. \cite[Chapter 29.1]{WEIN}.
If the supersymmetry generated by a superderivation  is unbroken, then 
the corresponding supercharge  is  given  as  a closable  linear operator 
on the GNS space for any  supersymmetric state 
by \eqref{eq:Qdef} in Proposition \eqref{prop:faithclo}. 
\end{rem}

\section{Supersymmetric fermion lattice systems}
\label{sec:SUSYfer}
\subsection{Fermion lattice systems on the CAR algebra}
\label{subsec:FERMIONlattice}
We shall formulate fermion lattice systems
on a  quasi-local $\cstar$-algebra.
 For simplicity
we will consider  the  $\nu$-dimensional cubic integer lattice  $\Znu$.  
However, it will be  clear that our setup is easily extended 
 to other lattices.  
For  $x=(x_i), \ y=(y_i) \in \Znu$,
let $|x-y|:=\displaystyle{\max_{1\le i \le \nu}} |x_i-y_i|$.
 For a  subset $\I$  of $\Znu$, 
  $|\I|$ denotes the volume, i.e. the number of sites
 in $\I$.  When  the volume of  $\I\subset\Znu$  is finite,
   we will  denote  $\I \Subset \Znu$. 
 For  each  $l \in  \NN \cup\{0\} \equiv\{0,1,2,3,\cdots\}$, 
  let us take   the following lattice cube  with its  edge length $l$
\begin{equation}
\label{eq:Cubl-def}
 \Cubl  := \{x=(x_1,\cdots, x_{\nu}) 
\in \Znu\ ;\  0 \le x_{i} \le l,  \ i=1, \cdots,\nu  \}.
\end{equation}
By definition its volume  $|\Cubl|$ is $(l+1)^{\nu}$.

We consider  interacting spinless fermions on $\Znu$.
  It is automatic   to extend the following formulation to    
 the case of  fermions  with  spins.
Let $\ai$ and $\aicr$ denote the annihilation operator 
and the creation operator of a spinless fermion at $i\in \Znu$, respectively. 
The canonical anticommutation relations (CARs) 
 are 
\begin{align}
\label{eq:CAR}
\{ \aicr, \aj \}&=\delta_{i,j}\, \unit, \nonumber \\
\{ \aicr, \ajcr \}&=\{ \ai, \aj \}=0.
\end{align}
For each $\I\Subset\Znu$, 
take the  finite-dimensional algebra $\FlI$
generated by $\{\aicr, \, \ai\, ;\;i\in \I\}$.
 It is isomorphic to ${\mathrm M}_{2^{|\I|}}(\CC)$, the algebra of all 
 $2^{|\I|}\times 2^{|\I|}$ complex matrices. 
 For  $\I \subset \J\Subset \Znu$, $\FlI$ is imbedded into $\FlJ$ as a 
subalgebra. Let
\begin{equation} 
\label{eq:CARloc}
\core:=\bigcup_{\I \Subset \Znu }\FlI.
\end{equation}
The total $\cstar$-system $\Fl$ is given by   
the  norm  completion of this   normed $\ast$-algebra $\core$.
  It  is  the CAR algebra.  
The dense $\ast$-subalgebra $\core$ in $\Fl$ is called the local algebra.
Let $\gamma$ denote the   automorphism on the $\cstar$-algebra 
 $\Fl$  determined  by    
 \begin{equation}
\label{eq:CARgamma}
\gamma(\ai)=-\ai, \quad \gamma(\aicr)=-\aicr,\quad i\in \Znu.
\end{equation}
This $\gamma$ gives a grading on $\Fl$:
\begin{equation}
\label{eq:CAREO}
 \Fle := \{F \in \Fl \; \bigl| \;   \gamma(F)=F  \},\quad  
 \Flo := \{F \in \Fl  \; \bigl| \;  \gamma(F)=-F  \},
 \end{equation}
\begin{equation}
\label{eq:CARbunkai}
\Fl=\Fle\oplus\Flo.
 \end{equation}
 For each  $\I\Subset\Znu$
\begin{equation}
\label{eq:CARIeo}
 \FlIe := \FlI\cap \Fle,\quad  \FlIo := \FlI\cap \Flo,
 \end{equation}
\begin{equation}
\label{eq:CARIbunkai-i}
\FlI=\FlIe\oplus\FlIo.
 \end{equation}
By  the CARs 
 \eqref{eq:CAR}  the following $\gamma$-locality holds:
\begin{equation}
\label{eq:glocality}
[A,\; B]_{\gamma}=0{\text {\ \ for every  \ }} A \in \FlI 
{\text {\ and \ }}B\in \FlJ {\text {\ \ if   \ }} \I\cap\J=\emptyset,\ \I,\J\Subset\Znu.
\end{equation}

\subsection{Superderivations  made by   local fermionic charges}
\label{subsec:FERMIONsuperderivation}
We shall   provide  a superderivation 
  by  `local fermionic charges' over  the lattice.
This  is analogous to  the well-known  construction  of a  time generator 
 by local Hamiltonians on fermion (or quantum spin) lattice systems, 
see \cite[Sect.6.2]{BR} \cite{AMrmp}.  
 Let  $\cha$ be a map  
from  the set of finite regions 
$\{\I; \; \I\Subset  \Znu\}$ to the local algebra $\core$ such that   
\begin{equation}
\label{eq:chadef}
\cha: \I \  \longmapsto \chaI\in \FlIo
\quad {\text{for every}}\ \I \Subset  \Znu.
\end{equation}
The above map $\cha$ will be called 
{\it{an assignment of local fermionic charges}} (over $\Znu$).
Suppose that $\cha$ and $\pot$ are 
two  assignments of local fermionic charges.
We can consider  their linear combination:
\begin{equation}
\label{eq:lin-assignment}
(c \cha +d \pot)(\I):=c \chaI+ d \potI, 
\quad c,d \in \CC \quad {\text{for every}}\ \I \Subset  \Znu.
\end{equation}
For any  $c,d \in \CC$,  $c\cha+d\pot$ is also  an assignment of local 
 fermionic charges.
The  {\it{conjugate}} of $\cha$ is defined  as 
\begin{equation}
\label{eq:chaastdef}
\chaast: \I \  
\longmapsto \chaI^{\ast}\in \FlIo
 \quad {\text{for every}}\ \I \Subset  \Znu.
\end{equation} 
If   $\cha=\chaast$, namely
\begin{equation}
\label{eq:chashermit}
\chaI=\chaI^{\ast} {\text {\ \ for every \ }} \I\Subset \Znu,
\end{equation}
then  $\cha$ is said to be  hermite.
 In the following $\cha$  is generically {\it{not}} hermite.

For a general assignment of local fermionic charges $\cha$, let 
\begin{equation}
\label{eq:chasdef}
\chasf:=\cha + \chaast,\quad \chass:=i (\cha - \chaast).
\end{equation} 
These  are  hermite by definition. Conversely 
\begin{equation}
\label{eq:chabychas}
\cha=\frac{1}{2}(\chasf-i\chass), \quad
 \chaast=\frac{1}{2}(\chasf+i\chass).
\end{equation}

We shall provide  assumptions upon $\cha$. 
 First assume that  
there exists an $r\ge 0$ such that  
\begin{equation}
\label{eq:range}
\chaI=0\quad {\text{whenever}}\ 
\diam(\I)\equiv \max_{x,y\in \I} |x-y|>r.
\end{equation}
The minimum non-negative  integer  $r$ satisfying 
 the condition \eqref{eq:range}  is  called the  range of $\cha$.
Obviously  $\chaast$ has the same 
  range $r$ as that of $\cha$.
Each of  $\chasf$ and  $\chass$ has its  finite range 
which is  equal to or less than $r$.

Let $\cha$ be an  assignment of  local fermionic charges
satisfying   the finite-range condition  \eqref{eq:range}.
Then  for every   $\I\Subset \Znu$ we can  define a  
$\gamma$-graded infinitesimal 
 transformation  on $\FlI$ as
\begin{equation}
\label{eq:DEFdelcha}
\delcha(A):= 
\sum_{ \X \cap \I \ne\emptyset,\, \X\Subset \Znu}  
[\cha(\X), \;  A]_{\gamma} {\text {\ \ for every \ }} A\in \FlI,
 \end{equation}
 where  the  summation is taken over  all finite subsets 
 $\{\X; \X\Subset\Znu\}$ 
that have  a  non trivial intersection with $\I$.
 We  may  suppress  `$\X\Subset \Znu$' for simplicity  
 when it is clear from  the context. 
  If  $\I \subset \J\Subset \Znu$, then it follows from  
the $\gamma$-locality 
\eqref{eq:glocality} 
 and the defining formula \eqref{eq:DEFdelcha} that for every  $A \in \FlI$
\begin{align*}
 \sum_{ \X \cap \J \ne\emptyset}  
[\cha(\X), \;  A]_{\gamma}
=\sum_{ \X \cap \I \ne\emptyset}  
[\cha(\X), \;  A]_{\gamma}+
 \sum_{ \X \cap \I=\emptyset,\,  \X \cap \J \ne\emptyset}  
[\cha(\X), \;  A]_{\gamma}
= \sum_{ \X \cap \I \ne\emptyset}  
[\cha(\X), \;  A]_{\gamma}.
 \end{align*}
Therefore   $\delcha$  is a well-defined linear map  on  $\core$, and the  
 set of  formulas  \eqref{eq:DEFdelcha} 
for $\{\I,\, \I\Subset \Znu\}$  uniquely yields  
\begin{align}
\label{eq:delcha-loc}
\delcha(A)&= \sum_{\X}
[\cha(\X), \; A]_{\gamma}\quad \text{for every}\  A \in \core.  
\end{align}
In the above formula   \eqref{eq:delcha-loc},  
 for each  $A\in \core$ only a finite 
 number of   $\{\X; \X\Subset \Znu\}$ give non-zero contributions.
We verify  that this $\delcha$ 
  satisfies the desiderata \eqref{eq:oddsuperder}  \eqref{eq:gleib}   
 for  superderivations.
Thus  the assignment  of local fermionic charges 
  provides   a  fermionic  generator.
 It is similar to  the  assignment  of local Hamiltonians that provides  
 a time-generator.

We note that the following nice property: 
\begin{equation}
\label{eq:delchaDALG}
\delcha(\core)\subset \core.
\end{equation}
More specifically,
\begin{equation}
\label{eq:delcha-range}
\delcha(\FlIeo)\subset\FlIroe {\text {\ \ for every\ \ }}
\I\Subset \Znu,
 \end{equation}
where  $\Ir\Subset\Znu$ denotes  the extended   finite region    
 of  $\I$  by the range  of $\cha$:
\begin{equation}
\label{eq:Ir}
\Ir:=\Bigl\{x\in \Znu \ ; \  \min_{y\in \I} |x-y| \le   r \Bigr\}. 
\end{equation}
  
The conjugate superderivation $\delchaast$  
 as in \eqref{eq:delastDEF} can be written by  
the conjugate  assignment  $\chaast$  of  \eqref{eq:chaastdef}.
Namely for each  $\I\Subset\Znu$
\begin{equation}
\label{eq:delchaast}
\delchaast(A)=\delchaastnaka(A)=
\sum_{ \X \cap \I \ne\emptyset}  
[\chaast(\X), \;  A]_{\gamma}{\text {\ \ for every \ }} A\in \FlI. 
 \end{equation}
Similarly  to  
\eqref{eq:delchaDALG}
  \eqref{eq:delcha-range} we have 
\begin{equation}
\label{eq:delchaastDALG}
\delchaast(\core)\subset \core,
\end{equation}
and 
\begin{equation}
\label{eq:delchaast-range}
\delchaast(\FlIeo)\subset\FlIroe {\text {\ \ for every\ \ }}
\I\Subset \Znu.
 \end{equation}

For the  superderivation $\delcha$ associated with $\cha$,
 which is  supposed to be  non-hermite,  
 the  pair of hermite  superderivations are given  as in \eqref{eq:dels-futatu}:\begin{equation}
\label{eq:delchas-def}
 \delchasf:=\delcha+\delchaast, \quad  
\delchass:=i(\delcha-\delchaast)\ \ \text{on}\  \core.
\end{equation}
As in \eqref{eq:del-gyaku}
\begin{equation}
\label{eq:delcha-gyaku}
\delcha=\frac{1}{2}(\delchasf-i\delchass), \quad
 \delchaast=\frac{1}{2}(\delchasf+i\delchass) \ \text{on}\  \core.
\end{equation}
 By \eqref{eq:chasdef} 
\eqref{eq:delchas-def}  for each $\I\Subset\Znu$
\begin{align}
\label{eq:delchas-lat}
\delchasf(A)&= 
\sum_{ \X \cap \I \ne\emptyset }  
[\chasf(\X), \;  A]_{\gamma}
=\sum_{ \X}  
[\chasf(\X), \;  A]_{\gamma}
\ {\text {\ \ for every \ }} A\in \FlI,
\nonum\\
\delchass(A)&= 
\sum_{ \X\cap \I \ne\emptyset}
[\chass(\X), \;  A]_{\gamma}
=\sum_{ \X}  
[\chass(\X), \;  A]_{\gamma}
\ {\text {\ \ for every \ }} A\in \FlI. 
\end{align}
We immediately  see that
\begin{equation}
\label{eq:delchasDALG}
\delchasf(\core)\subset \core,\quad
\delchass(\core)\subset \core,
\end{equation}
and more specifically,
\begin{equation}
\label{eq:delchas-range}
\delchasf(\FlIeo)\subset\FlIroe,\quad 
\delchass(\FlIeo)\subset\FlIroe
 {\text {\ \ for every\ \ }} \I\Subset \Znu.
\end{equation}

Second, assume  that  $\cha$ is uniformly bounded,  
\begin{equation} 
\label{eq:chanorm}
\|\cha \|_{\infty} := 
\sup _{\I  \Subset  \Znu} 
\|\chaI \|<\infty.
\end{equation}
Then due to  \eqref{eq:chaastdef}  
\begin{equation}
\label{eq:chaastnorm}
\| \chaast \|_{\infty}  = \|  \cha \|_{\infty}<\infty.
\end{equation}
The uniform  boundedness 
of $\cha$ and $\chaast$
 together with \eqref{eq:chasdef} implies
 the  uniform  boundedness 
 of  $\chasf$ and $\chass$:
\begin{equation}
\label{eq:chasnorm}
\|\chasf \|_{\infty}  \le  2 \|  \cha \|_{\infty}<\infty,\quad 
\|\chass \|_{\infty}  \le  2 \|  \cha \|_{\infty}<\infty.
\end{equation}

Third,  assume that 
the superderivation  $\delcha$ associated with $\cha$
 satisfies the  nilpotent  condition:
\begin{equation}
\label{eq:nil-delcha}
\delcha\cdot\delcha=\zeromap  
{\text {\ \ on\ \ }} \core. 
\end{equation}
 We may say
 that $\cha$ is `nilpotent' if
 its   associated superderivation $\delcha$  
 is  nilpotent  \eqref{eq:nil-delcha}. 
It is a strange  terminology, but may be 
  a convenient shorthand.

Let us summarize the terminologies given  so far.
\begin{defn}
\label{defn:cha}
A  map $\cha:\I \ \longmapsto \chaI\in \FlIo$
 defined  on $\{\I; \; \I \Subset  \Znu \}$  is  called  an  
assignment of  local fermionic charges on the  fermion lattice system.
If  $\cha$ satisfies the condition \eqref{eq:range} with some finite $r$, 
 then it is said to be  finite range. 
For a finite-range   assignment of  local fermion charges
  $\cha$, the linear map  $\delcha$ from 
 the local algebra $\core$ into $\core$ by the formula  \eqref{eq:DEFdelcha}
 is called  the superderivation associated with $\cha$.
The set of  all bounded  finite-range assignments of 
 local fermion charges is denoted  by $\Chasp$.
The set of all $\cha\in \Chasp$ whose associated superderivation  
$\delcha$  satisfies the  nilpotent condition  \eqref{eq:nil-delcha} 
is denoted  by $\Chaspn$.
\end{defn}

\begin{rem} 
\label{rem:cha-trans-cov}
We do not assume  translation 
 covariance  for $\cha$ in  Definition \eqref{defn:cha}.
Actually  we will consider   an example which is  
  periodic, but not translationally  covariant.
\end{rem}

\begin{rem} 
\label{rem:nontrivial-nilp}
To make  a concrete  example  of $\Chaspn$, 
the nilpotent condition \eqref{eq:nil-delcha}
is  the most  non trivial requirement.
Later  we  will show   a  criterion of the nilpotent condition
 which  can be easily checked for  $\cha\in \Chasp$.
\end{rem}

\begin{rem} 
\label{rem:Csplinear}
It is obvious that  $\Chasp$ is a $\CC$-linear space. 
 However, $\Chaspn$ is not a linear space,
 since the  nilpotent condition \eqref{eq:nil-delcha}
 is not generally preserved under linear summation.
\end{rem}

The following is due to Definition 
  \eqref{defn:unbroken--broken}.
\begin{defn}
\label{defn:chaunbroken}
Let $\cha$ be an element of $\Chaspn$
   given in  Definition \eqref{defn:cha}.
If there exists a supersymmetric state $\vp$
 with respect to the associated superderivation $\delcha$,
 then it is said that 
$\cha\in \Chaspn$ gives an  unbroken-supersymmetry model,
 or  in short $\cha\in \Chaspn$ is  unbroken supersymmetry.
The set of all $\cha\in \Chaspn$  
giving  an  unbroken-supersymmetry model 
 is denoted  by $\Chaspnun$.
\end{defn}

\subsection{Supersymmetry  formula on the local algebra}
\label{subsec:derhcha}
The following lemma is obvious.
\begin{lem}
\label{lem:ast-nil}
Suppose that a finite-range assignment of local fermion 
charges $\cha$ is given.
If the associated superderivation  $\delcha$
 is nilpotent, namely  the condition \eqref{eq:nil-delcha} is satisfied,
  then the conjugate  superderivation 
$\delchaast=\delchaastnaka$  is also  nilpotent{\rm{:}}
\begin{equation}
\label{eq:nil-delchaast}
\delchaast\cdot\delchaast=\zeromap  
{\text {\ \ on\ \ }} \core. 
\end{equation}
\end{lem}
\begin{proof}
By noting  \eqref{eq:delastDEF}
 \eqref{eq:oddsuperder}
 we see  that  Eq.\eqref{eq:nil-delcha}
 and  Eq.\eqref{eq:nil-delchaast} are equivalent.
\end{proof}

\begin{rem} 
\label{rem:ast-closed}
It is quite obvious  that  $\Chasp$  is closed under the $\ast$-operation 
 by \eqref{eq:chaastdef}.
By Lemma \eqref{lem:ast-nil} 
 $\Chaspn$ is also closed under the $\ast$-operation.
\end{rem}

We {\it{define}} a derivation which has the  required supersymmetric form.
\begin{defn}
\label{defn:derhcha}
For any  $\cha\in \Chaspn$  define the following derivation  
\begin{equation}
\label{eq:def-derhcha}
\Derhcha:=
\delchaast\cdot\delcha+ \delcha\cdot\delchaast
{\text {\ \ on\ \ }}\core.
\end{equation}
\end{defn}
Since  both   $\delchaast\cdot\delcha$ and $\delcha\cdot\delchaast$
 can be defined on $\core$ due to 
\eqref{eq:delchaDALG} \eqref{eq:delchaastDALG},
 the above definition makes sense.
Furthermore 
\begin{equation}
\label{eq:DerchaDALG}
\Derhcha(\core)\subset \core,
\end{equation}
more specifically,  
due to  \eqref{eq:delcha-range} \eqref{eq:delchaast-range}  
\begin{equation}
\label{eq:Derhcha-range}
\Derhcha(\FlIeo)\subset\FlItworeo {\text {\ \ for every\ \ }}
\I\Subset \Znu.
 \end{equation}
It is easy to see that  $\Derhcha$   
 is a derivation  (satisfying   the Leibniz rule).
From  the oddness  of  superderivations 
 \eqref{eq:oddsuperder} and 
the definition  of $\Derhcha$ it follows that  
\begin{equation}
\label{eq:Derhcha-even}
\Derhcha\cdot \gamma = \gamma \cdot \Derhcha
{\text {\ \ on\ \ }} \core.
\end{equation}

It is possible to  rewrite  the infinitesimal supersymmetry formula 
 of  Definition \eqref{defn:derhcha} 
 in terms of  the pair of (independent) hermite superderivations
 $\delchasf$ and $\delchass$. 
\begin{prop}
\label{prop:susy-cha-sym}
The set of relations \eqref{eq:nil-delcha} and 
\eqref{eq:def-derhcha} is equivalent to the following 
set of  relations  in terms of the hermite superderivations 
$\delchasf$ and $\delchass$ in  Eq.\eqref{eq:delchas-def}{\rm{:}}
\begin{equation}
\label{eq:delchas-independ}
\delchasf\cdot\delchass+\delchass\cdot \delchasf=\zeromap
{\text {\ \ on\ \ }}
\core,  
\end{equation}
\begin{equation}
\label{eq:cha-susysym}
\Derhcha={\delchasf}^{\!\!\!2}={\delchass}^{\!\!\!2}
{\text {\ \ on\ \ }}\core.  
\end{equation}
\end{prop}

\begin{proof}
First we will derive  the  formulas  
\eqref{eq:delchas-independ}  \eqref{eq:cha-susysym} from  
 \eqref{eq:nil-delcha} and \eqref{eq:def-derhcha}.
By noting \eqref{eq:delchas-def} and Lemma \eqref{lem:ast-nil}
 we have 
\begin{align*}
\delchasf\cdot\delchass&=i(\delcha\cdot\delcha- \delchaast\cdot\delchaast)
+i(\delchaast\cdot\delcha-\delcha\cdot\delchaast) \\
&=\zeromap+i(\delchaast\cdot\delcha-\delcha\cdot\delchaast) \\
&=i(\delchaast\cdot\delcha-\delcha\cdot\delchaast),
\end{align*} 
and 
\begin{align*}
\delchass\cdot\delchasf&=i(\delcha\cdot\delcha- \delchaast\cdot\delchaast)
-i(\delchaast\cdot\delcha-\delcha\cdot\delchaast) \\
&=\zeromap-i(\delchaast\cdot\delcha-\delcha\cdot\delchaast) \\
&=-i(\delchaast\cdot\delcha-\delcha\cdot\delchaast).
\end{align*} 
These yield 
\begin{equation*}
\delchasf\cdot\delchass+\delchass\cdot\delchasf=\zeromap.
\end{equation*} 
We have 
\begin{equation*}
\label{eq:}
\delchasf\cdot\delchasf=
(\delchaast\cdot\delcha+\delcha\cdot\delchaast)+
\delcha\cdot\delcha+ \delchaast\cdot\delchaast=\Derhcha+\zeromap
+\zeromap=\Derhcha,  
\end{equation*} 
and similarly 
\begin{equation*}
\delchass\cdot\delchass=\Derhcha. 
\end{equation*} 

We will show   the converse direction.  
Assume  that Eqs.\eqref{eq:delchas-independ} and  \eqref{eq:cha-susysym} 
 are satisfied. 
By noting \eqref{eq:delcha-gyaku} we have 
\begin{equation*}
\delcha\cdot\delcha=\frac{1}{4}(\delchasf\cdot\delchasf-
\delchass\cdot\delchass) -\frac{i}{4}
(\delchasf\cdot\delchass + \delchass\cdot\delchasf)
=\frac{1}{4}(\Derhcha-\Derhcha) +\zeromap=\zeromap.
\end{equation*} 
We have  
\begin{equation*}
\label{eq:}
\delchaast\cdot\delcha+\delcha\cdot\delchaast  
=\frac{1}{4}( \delchasf\cdot\delchasf+ \delchass\cdot\delchass
)\cdot 2 =
\frac{1}{2}(\Derhcha+ \Derhcha) =\Derhcha.
\end{equation*}
Thus we have shown the assertion.  
\end{proof}

At  this stage, the derivation $\Derh$ 
defined in  Definition \eqref{defn:derhcha} is not  related to 
 time evolution.  
We have not yet obtained  a  supersymmetric dynamics which  
 will be shown in the next subsection.
(A similar   problem  arises     
 to  formulate supersymmetry 
in a  fermion-boson $\cstar$-system.  See  
\cite{BUGR}  \cite{MORIYAnew} for the detail.)

\subsection{Supersymmetric $\cstar$-dynamics on the CAR algebra}
\label{subsec:C-sta-dyn}
The supersymmetry formula \eqref{eq:def-derhcha} in  Definition 
\eqref{defn:derhcha}  made by  a 
 nilpotent  superderivation \eqref{eq:nil-delcha}  expresses the same  
 supersymmetry algebra  as described in Sect.\ref{sec:overview}.
 However,  to make a  more complete form 
 global time evolution should be involved  as well as  its infinitesimal 
 generator.
We immediately see that there exists a strongly continuous one parameter
group of $\ast$-automorphisms generated by the derivation
 $\Derhcha$ on the CAR algebra $\Fl$, 
 since  $\Derhcha$ defined on the local algebra  is of finite range as 
noted in \eqref{eq:Derhcha-range}. 
 We shall  provide  its detailed construction  
 and  discuss 
 its characteristic properties due to  hidden supersymmetry.

\subsubsection{Expansion of the iteration  
 of superderivations}
\label{subsubsec:ingredient}
 Take  any nilpotent  finite-range  assignment of local fermion 
charges $\cha\in \Chaspn$ of   Definition {\rm{\eqref{defn:cha}}}.
We will denote  the  hermite  assignments of  local fermion charges
 $\chasf$  and $\chass$ in \eqref{eq:chasdef} simply  by  $\chas$.
 There will  arise  no  essential difference  between  
$\chasf$  and $\chass$  in the following.
 Of course  we have to use `$1$' and  `$2$' consistently.

Consider  $\delchasn\equiv \underbrace{\delchas \cdot \delchas\cdot\, \cdots 
\,\cdot \delchas}_{n \;{\rm{times}}}$ for  $n\in \NN$,  
i.e.  the $n$th  iterate of the map  $\delchas$.
Due to the finite-range condition  \eqref{eq:delchas-range} 
for $\delchas$,
 $\delchasn$ is well-defined and  finite range for each $n\in \NN$. 
Particularly, for $n=2m-1$ ($m\in \NN$)
$\delchasn$ is a superderivation 
from  $\core$ into $\core$, while  for $n=2m$ ($m\in \NN$)
it is a derivation from  $\core$ into $\core$.
Take any  $\I\Subset\Znu$.  
 For every $A\in \FlI$ 
 by using  the formula \eqref{eq:delchas-lat} repeatedly 
we have 
\begin{align}
\label{eq:delchasn-I}
&\ \delchasn(A) \nonum\\
&=
\sum_{\X_1, \X_2, \cdots, \X_n}
\Bigl[\chas(\X_n)
, \;\cdots {\Bigl[} \chas(\X_2),\;  \Bigl[\chas(\X_1), \;  A\Bigr]_{\gamma} 
{\Bigr]}_{\gamma} \cdots \Bigl]_{\gamma}\nonum\\
&=\!\!\!\!\!\!\!\!\!\!\!\!\!\!\!\!\!\!\!
\sum_{ \bigl\{ \left(\X_1, \X_2, \cdots,  \X_n\right)\,|\,
\X_1  \cap \V_{0}\ne \emptyset,\   
\X_2  \cap \V_{1}\ne\emptyset, \ \cdots\ {\text{and}}\  
\X_n \cap \V_{n-1}\ne \emptyset \bigr\}}
\!\!\!\!\!\!\!\!\!\!\!\!\!\!\!\!\!\!\!\!\!\!\!\!\!\!\!\!\!\!\!\!\!\!\!\!\!\!\!\!\! 
\Bigl[\chas(\X_n)
, \;\cdots {\Bigl[} \chas(\X_2),\;  \Bigl[\chas(\X_1), \;  A\Bigr]_{\gamma} 
{\Bigr]}_{\gamma} \cdots \Bigl]_{\gamma}\nonum\\
& \quad \in \FlInr,
\end{align}
 where we have defined
\begin{equation}
\label{eq:Vj}
\V_{0}:=\I,\quad  
\V_{j}:=\X_j \cup \X_{j-1}\cup \cdots \cup \X_1 \cup \Vzero
\ {\text{for}}\ j \in \{1,2,3,\cdots \}.   
\end{equation}
The condition upon  the  multiplets  
$\left(\X_1, \X_2, \cdots \cdots,  \X_n\right)$ 
 in the second  equality of   
\eqref{eq:delchasn-I} is due to the  $\gamma$-locality \eqref{eq:glocality}; 
the graded commutator  of  $\chas(\X)$ with any element outside of $\X$  
 vanishes, and  
 $\delchasn(A)\in \FlInr$ is due to  \eqref{eq:delchas-range}.
Hence the finite regions  $\Vj$ ($j \in \{1,2,3,\cdots \}$) 
which  may  give non-zero contribution  
  to   the formula \eqref{eq:delchasn-I} should   satisfy    
\begin{equation}
\label{eq:Vjvol}
\Vj\subset \Ijr,\  \
|\Vj|  \le  |\I| + j (|\Cubr|-1) < |\I|+j \cdot (r+1)^{\nu}.
\end{equation}

We   will consider   the derivation  
$\Derhcha$ of  Definition \eqref{defn:derhcha}.
By  the identity  $\Derhcha={\delchas}^{\!\!2}$ 
 on $\core$ given in  Proposition \eqref{prop:susy-cha-sym}
  Eq.\eqref{eq:delchasn-I} gives  
 a similar expansion  formula
for the $n$th iteration  of $\Derhcha$  ($n\in \NN$): 
 For each   $A\in \FlI$  
\begin{align}
\label{eq:derhn}
&\ \Derhchan(A)={\delchas}^{\!\!2n}(A)\nonum\\
&=\!\!\!\!\!\!\!\!\!\!\!\!\!\!\!\!\!\!\!\!
\sum_{ \bigl\{ \left(\X_1, \X_2, \cdots,  \X_{2n}\right)\,|\,
\X_1  \cap \Vzero\ne \emptyset,\   
\X_2  \cap \Vone \ne\emptyset, \cdots\ {\text{and}}\  
\X_{2n} \cap \V_{2n-1}\ne \emptyset \bigr\}}
\!\!\!\!\!\!\!\!
\!\!\!\!\!\!\!\!\!\!\!\!\!\!\!\!\!\!\!\!\!\!\!\!\!\!\!\!\!\!\!\!\!
\Bigl[\chas(\X_{2n})
, \;\cdots {\Bigl[} \chas(\X_2),\;  \Bigl[\chas(\X_1), 
\;  A\Bigr]_{\gamma} {\Bigr]}_{\gamma} \cdots \Bigl]_{\gamma}\nonum\\
&\quad \in \FlItwonr,
\end{align}
where  the  notation \eqref{eq:Vj} is used.

Let us introduce   a set of  bounded superderivations 
indexed by  $\Lam \Subset \Znu$ as  
\begin{equation}
\label{eq:delchaslam}
\delchaslam(F):= \sum_{ \X \cap \Lam \ne\emptyset } 
[\chas(\X), \;  F]_{\gamma}
\ {\text {\ \ for every    \ }} F  \in \Fl.
 \end{equation}
As the range of $\chas$
 is equal to or less than $r$, for each $\Lam$ fixed, 
  the finite subsets $\X \Subset\Znu$ which  may  give  non-zero contribution  
  to  the  formula \eqref{eq:delchaslam}  are all included in $\Lamr$.
If  $\I \subset \Lam\Subset \Znu$, then 
 \begin{equation*}
 \delchas(A)=\delchaslam(A)  {\text {\ \ for  every  \ }}  A \in \FlI.
 \end{equation*}
Hence 
\begin{equation}
\delchas(A)=\Lamlim \delchaslam(A) 
{\text {\ \ for  each  \ }}  A \in \core, 
\end{equation}
where   `$\mugen$'  means  that the  net 
 $\{\Lam ;\; \Lam\Subset \Znu \}$  tends  to the whole $\Znu$, 
 i.e. it eventually contains   any  finite subset of $\Znu$.

For each $\Lam\Subset \Znu$
 define  the bounded derivation associated with $\Lam$ 
 by
\begin{equation}
\label{eq:derhchalam}
\Derhchalam(F):= \delchaslam \cdot \delchaslam (F) 
{\text {\ \ for every  \ }} F  \in \Fl.
 \end{equation}
Note  that the above  $\Derhchalam$ actually depends on the choice 
 of $\chasf$ and $\chass$.  
 However, this choice does not change the argument in what  follows.  
By \eqref{eq:delchaslam} it is  written in terms of
$\chas$ as 
\begin{equation}
\label{eq:derhchalam-sum}
\Derhchalam(F)=
\sum_{\X_1\cap \Lam \ne\emptyset,\; \X_2\cap \Lam \ne\emptyset}
\Bigl[ \chas(\X_2),\;  \Bigl[\chas(\X_1), \;  F\Bigr]_{\gamma}
\Bigr]_{\gamma}.
\end{equation}
Let us consider  its  $n$th iterate map ($n\in\NN$), i.e.    
$\Derhchalamn\equiv 
\underbrace{\Derhchalam  \cdot \Derhchalam \cdots 
\Derhchalam}_{n \;{\rm{times}}}$, which 
  is obviously a  bounded derivation of $\Fl$.
 For every  $A\in \FlI$ with a fixed  $\I\Subset\Znu$ we have 
\begin{align}
\label{eq:derhchalamn}
 &\Derhchalamn(A)\!\!\! \nonum\\
&=\!\!\!\!\!\!\!\!\!\!\!\!
\sum_{\X_1\cap \Lam \ne\emptyset,\; 
\X_2\cap \Lam \ne\emptyset, \cdots ,\X_{2n}\cap \Lam \ne\emptyset}
\!\!\!\!\!\!\!\!\!\!\!\!\!\!\!\!\!\!\!\!\!\!
\Bigl[\chas(\X_{2n})
, \;\cdots {\Bigl[} \chas(\X_2),\;  \Bigl[\chas(\X_1), \;  A\Bigr]_{\gamma} 
{\Bigr]}_{\gamma} \cdots \Bigl]_{\gamma}
\nonum  \\
&= 
\!\!\!\!\!\!\!\!\!\!\!\!\!\!\!\!\!\!\!\!
\sum_{ \bigl\{ \left(\X_1, \X_2, \cdots,  \X_{2n}\right)\,|\,
\X_i  \cap \V_{i-1}\ne \emptyset 
\ {\text{and}}\ \X_i\cap\Lam\ne \emptyset \ {\text{for each}}\ i \in \{1,2,\cdots, 2n\} 
\bigr\}} \!\!\!\!\!\!\!\!\!\!\!\!\!\!\! \!\!\!\!\!\!\!\!\!\!\!\!\!\!\!\!\!\!\!\!\!\!\!\!\!\!\!\!\!\!\!\!\!\!\!\!\!\!\!\!\!\!\!\!\!\!
\Bigl[\chas(\X_{2n}), \cdots \!\!\ {\Bigl[} \chas(\X_2),\;  \Bigl[\chas(\X_1), \; A\Bigr]_{\gamma} 
{\Bigr]}_{\gamma} \cdots \Bigl]_{\gamma} \nonum\\
&\quad \in  \Fl(\I\cup \Lamr)\cap \Fl(\Itwonr),
\end{align}
where  the  notation  \eqref{eq:Vj} is used again.
By comparing  Eq.\eqref{eq:derhchalamn} with Eq.\eqref{eq:derhn} 
there exists  a sufficiently large $\Lam_\circ\Subset \Znu$ 
 which depends on the given $n\in \NN$ and  $\I \Subset \Znu$
such  that for any finite $\Lam\supset \Lam_{\circ}$
\begin{equation}
\label{eq:lamn-large}
 \Derhchalamk(A)=
\Derhchak(A),\ \ k=1,\;2,\; \cdots\;,n-1, \;n,
 {\text {\ \ for  every  \ }} A  \in \FlI.
\end{equation}
For each $A\in \core$ and $n\in \NN$ 
\begin{equation}
\label{eq:vlimderhlam}
\Lamlim \Derhchalamk(A)= \Derhchak(A),\ \ k=1,\;2,\; \cdots\;,n-1, \;n.
\end{equation}

We can   rewrite the above expansion formulas of 
 $\Derhchalam$ and  $\Derhcha$  
 in terms of $\chas$ as follows.
\begin{lem}
\label{lem:pair}
Let $\chas$ denote $\chasf$ or $\chass$ given  in  \eqref{eq:chasdef}. 
Let $\Derhchalam$ {\rm{(}}$\Lam \Subset \Znu${\rm{)}} denote 
the   bounded derivation of $\Fl$ defined in \eqref{eq:derhchalam}.
Let $\Derhcha$ denote the derivation on  $\core$ 
 as in \eqref{eq:cha-susysym}.
Then  for each $F \in \Fl$, 
\begin{align}
\label{eq:derhchalam-pair}
&\quad \Derhchalam(F)\nonum\\
&=\frac{1}{2}
 \sum_{\substack{  \X_1\Subset \Znu,\,\X_2\Subset \Znu
\\ \X_1\cap \Lam \ne\emptyset,\, \X_2\cap \Lam \ne\emptyset \\
 \X_1\cap \X_2 \ne\emptyset}}
\Bigl[  \bigl\{\chas(\X_2),\;   \chas(\X_1)\bigr\}, \; F \Bigl]
= \sum_{\substack{  \X_1\Subset \Znu, \, \X_2\Subset \Znu
\\ \X_1\cap \Lam \ne\emptyset,\, \X_2\cap \Lam \ne\emptyset \\
 \X_1\cap \X_2 \ne\emptyset}} 
\Bigl[ \chas(\X_2) \chas(\X_1), \;  F \Bigr].
\end{align}
For each $A\in \core$, 
\begin{align}
\label{eq:derhcha-pair}
\Derhcha(A)
=\frac{1}{2}
 \sum_{\substack{\X_1\Subset \Znu, \, \X_2\Subset \Znu
\\  \X_1\cap \X_2 \ne\emptyset}}
\Bigl[  \bigl\{\chas(\X_2),\;   \chas(\X_1)\bigr\}, \; A \Bigl]
=\sum_{\substack{\X_1\Subset \Znu, \, \X_2\Subset \Znu
\\ \X_1\cap \X_2 \ne\emptyset}}
\Bigl[ \chas(\X_2) \chas(\X_1),
\;  A \Bigr],
\end{align}
where the summation is a finite sum.
\end{lem}

\begin{proof}
For a given $F\in\Fl$ we have its unique 
 decomposition  $F=\Fp+\Fm$, where  
$\Fp=\frac{1}{2}(F+\gamma(F))\in \Fle$ and  $\Fm
=\frac{1}{2}(F-\gamma(F))\in \Flo$. From \eqref{eq:derhchalam-sum} we have
\begin{align*}
&\Derhchalam(F)\nonum\\
 &= \Derhchalam(\Fp)+ \Derhchalam(\Fm) \nonum\\
&=\!\!\!\!\!\!\sum_{\X_1\cap \Lam \ne\emptyset,\; \X_2\cap \Lam \ne\emptyset}
\!\!\!\!\!\!\Bigl\{ \chas(\X_2),\;  \ \chas(\X_1)\Fp-   \Fp \chas(\X_1) \Bigr\}
+\Bigl[ \chas(\X_2),\;  \chas(\X_1)\Fm +\Fm \chas(\X_1)
\Bigr]  \nonum \\
&=\!\!\!\!\!\!\!\!\!\!\!\!\sum_{\X_1\cap \Lam \ne\emptyset,\; \X_2\cap \Lam \ne\emptyset}\!\!\!\!\!\!\!\!\!\!\!\!
\chas(\X_2)  \chas(\X_1)\Fp 
+ \chas(\X_1) \Fp \chas(\X_2)
 -  \chas(\X_2)  \Fp \chas(\X_1) -\Fp \chas(\X_1) \chas(\X_2) \nonum\\
&+\!\!\!\!\!\!\!\!\!\!\!\!\!\!\!\sum_{\X_1\cap \Lam \ne\emptyset,\; \X_2\cap \Lam \ne\emptyset}\!\!\!\!\!\!\!\!\!\!\!\!
 \chas(\X_2)  \chas(\X_1)\Fm 
- \chas(\X_1) \Fm  \chas(\X_2)+  \chas(\X_2)  \Fm \chas(\X_1)  
 -\Fm \chas(\X_1) \chas(\X_2) \nonum \\
&=\!\!\!\!\!\!\!\!\!\!\!\!\sum_{\X_1\cap \Lam \ne\emptyset,\; \X_2\cap \Lam \ne\emptyset}\!\!\!\!\!\!\!\!\!\!\!\!
\chas(\X_2)  \chas(\X_1)\Fp 
+ \Bigl(\chas(\X_1) \Fp \chas(\X_2)
 -  \chas(\X_2)  \Fp \chas(\X_1)\Bigr) -\Fp \chas(\X_1) \chas(\X_2) \nonum\\
&+\!\!\!\!\!\!\!\!\!\!\!\!\!\!\!\sum_{\X_1\cap \Lam \ne\emptyset,\; \X_2\cap \Lam \ne\emptyset}\!\!\!\!\!\!\!\!\!\!\!\!
 \chas(\X_2)  \chas(\X_1)\Fm +\Bigl(
- \chas(\X_1) \Fm  \chas(\X_2)+  \chas(\X_2)  \Fm \chas(\X_1)\Bigr)  
 -\Fm \chas(\X_1) \chas(\X_2) \nonum \\
&= \!\!\!\!\!\!\!\!\!\!\!\!\!\!\!
\sum_{\X_1\cap \Lam \ne\emptyset,\; \X_2\cap \Lam \ne\emptyset}
\!\!\!\!\!\!\!\!\!
\chas(\X_2)  \chas(\X_1)\Fp 
 -\Fp \chas(\X_1) \chas(\X_2)
 +\chas(\X_2)  \chas(\X_1)\Fm 
 -\Fm \chas(\X_1) \chas(\X_2)\nonum \\
&= \!\!\!\!\!\!\!\!\!\!\!\!\!\!\!
\sum_{\X_1\cap \Lam \ne\emptyset,\; \X_2\cap \Lam \ne\emptyset}
\!\!\!\!\!\!\!\!\!
\chas(\X_2)  \chas(\X_1)\Fp 
 -\Fp \chas(\X_2) \chas(\X_1)
 +\chas(\X_2)  \chas(\X_1)\Fm 
 -\Fm \chas(\X_2) \chas(\X_1)\nonum \\
&=\sum_{\X_1\cap \Lam \ne\emptyset,\; \X_2\cap \Lam \ne\emptyset}
\Bigl[  \chas(\X_2)  \chas(\X_1),\; \Fp+\Fm \Bigl] 
=\sum_{\X_1\cap \Lam \ne\emptyset,\; \X_2\cap \Lam \ne\emptyset}
\Bigl[  \chas(\X_2)  \chas(\X_1),\; F \Bigl]  
\nonum\\
&= \frac{1}{2}
 \sum_{
\X_1\cap \Lam \ne\emptyset,\; \X_2\cap \Lam \ne\emptyset}
\!\!\!\!
\Bigl[\chas(\X_2)  \chas(\X_1) + \chas(\X_1) \chas(\X_2),\; F \Bigr]\nonum\\
&=\frac{1}{2}
 \sum_{\X_1\cap \Lam \ne\emptyset,\; \X_2\cap \Lam \ne\emptyset}
\!\!\!\!
\Bigl[  \bigl\{\chas(\X_2),\;   \chas(\X_1)\bigr\}, \; F \Bigl]. 
\end{align*}
By noting the vanishing anti-commutator  
$\left\{\chas(\X_2),\;\chas(\X_1)\right\}=0$ if 
$\X_1\cap\X_2= \emptyset$ we get   
\begin{align*}
\Derhchalam(F)
&= \frac{1}{2}
 \sum_{\X_1\cap \Lam \ne\emptyset,\; \X_2\cap \Lam \ne\emptyset\atop{
 \X_1\cap \X_2 \ne\emptyset}}
\Bigl[  \bigl\{\chas(\X_2),\;   \chas(\X_1)\bigr\}, \; F \Bigl]\nonum\\
&= \frac{1}{2}
 \sum_{\X_1\cap \Lam \ne\emptyset,\; \X_2\cap \Lam \ne\emptyset\atop{
 \X_1\cap \X_2 \ne\emptyset}}
\Bigl[  \chas(\X_2)\chas(\X_1)
+ \chas(\X_1)\chas(\X_2), \; F \Bigl] 
\nonum\\
&= \sum_{\X_1\cap \Lam \ne\emptyset,\; \X_2\cap \Lam \ne\emptyset\atop{
 \X_1\cap \X_2 \ne\emptyset}}
\Bigl[  \chas(\X_2)\chas(\X_1), \; F \Bigl].
\end{align*}
Hence   Eq.\eqref{eq:derhchalam-pair} is shown.  
We obtain  \eqref{eq:derhcha-pair} from  \eqref{eq:derhchalam-pair} 
 for any $A\in\core$ due to  the asymptotic formula 
\eqref{eq:vlimderhlam}  by  taking  $\mugen$.
Let $\I\Subset\Znu$  be  the least subset such that $A\in \FlI$.
As $\chas(\X_2)\chas(\X_1)\in \coree$, 
 if $\X_1 \cup \X_2$ is included in $\Ic$ (i.e. the complement of $\I$), 
 then the commutator 
 for such $\X_1$ and $X_2$  appeared in \eqref{eq:derhcha-pair}
vanishes.
 Hence Eq.\eqref{eq:derhcha-pair} is actually  a finite sum. 
\end{proof}

\begin{rem} 
\label{rem:pair}
In  Lemma \eqref{lem:pair} the finite-range  assignment of local 
fermion charges $\cha$ needs not to be nilpotent.
This fact will be exploited later to show   
 Proposition \eqref{prop:delcha-nil}. 
\end{rem} 

\subsubsection{Global time evolution}
\label{subsubsec:time}
We will construct a global time evolution from 
  $\cha\in \Chaspn$ of   Definition {\rm{\eqref{defn:cha}}}.
For each $\Lam\Subset \Znu$ by taking the exponential 
 of  the bounded derivation $\Derhchalam$
let 
\begin{equation}
\label{eq:altlam}
\altlam(F):= \exp({it \Derhchalam})(F)  
{\text {\ \ for  every  \ }} t\in \R{\text {\ \ and  \ }}   F  \in \Fl.
 \end{equation}
By  \eqref{eq:derhchalamn}  for any $\I\Subset\Znu$
\begin{equation}
\label{eq:altlam-affil}
\altlam(A)\in \Fl(\I\cup \Lamr)
{\text {\ \ for  every  \ }} t\in \R{\text {\ \ and  \ }}   
 A\in \Fl(\I). 
\end{equation}
We  note its  expansion formula:
\begin{equation}
\label{eq:Fnlam}
\altlam(F)=\sum_{n=0}^{\infty}\frac{(it)^{n}}{n!}\Derhchalamn(F) 
{\text {\ \ for  every  \ }} t\in \R{\text {\ \ and  \ }}   F  \in \Fl.
 \end{equation}
We will get a  time evolution on the whole system $\Fl$  
 by taking  infinite volume limit 
 of the  inner  one-parameter group of 
$\ast$-automorphisms $\{\altlam\,;\;t\in \R\}$ of $\Fl$.
It has been known that 
  the generator $\Derhcha$  of a finite-range interaction  
is analytic on the local algebra $\core$, see \cite{ROB68} and 
 Sect.7.3 of \cite{STRbook}.
We shall  recapture its proof in the following lemma.
\begin{lem}
\label{lem:multiple-est}
Let  $\cha\in\Chaspn$ of Definition {\rm{\eqref{defn:cha}}} and 
 let $r$ denote the range of  $\cha$.
Let $\chas$ denote $\chasf$ or  $\chass$   
 given in   \eqref{eq:chasdef}. 
Let $\Derhcha$ denote the derivation defined on   $\core$ 
  given in  \eqref{eq:def-derhcha}
in  Definition {\rm{\eqref{defn:derhcha}}}.
For each $\Lam\Subset \Znu$
let $\Derhchalam$ denote the bounded derivation of $\Fl$
given in \eqref{eq:derhchalam}.
Then for each $A\in \core$ and $n\in \NN$, 
\begin{align}
\label{eq:derhnA-est}
\frac{1}{n!}\bigl\| \Derhchan(A)\bigr\| \le \kchas \cdot \Mchasn
\end{align}
and 
\begin{align}
\label{eq:derhlamnA-est}
\frac{1}{n!}\bigl\|\Derhchalamn(A)\bigr\| \le \kchas\cdot\Mchasn,
\end{align}
where  $\I\Subset \Znu$ denotes   the least subset 
 such that   $A\in \Fl(\I)$, and  
\begin{align}
\label{eq:Lchas}
\kchas&:= \|A \| \cdot 
\exp\Bigr(4 \bigl\{|\I|-2(r+1)^{\nu}\bigr\} \Lchas \Bigr), \quad 
\Mchas:=\exp\Bigr(8 (r+1)^{\nu}\Lchas\Bigl),\nonum\\
\Lchas&:= \sup _{i  \in \Znu}\sum_{\Y;\,\Y\cap\X\ne\emptyset}
\; \; \sum_{\X;\,\X\ni i} \| \chas(\Y) \| \cdot\| \chas(\X) \|<\infty.
\end{align}
(In the above  defining   formula of $\Lchas$, 
first fix  $i\in\Znu$, second  
 take a finite subset $\X$ that contains  the fixed site $i$,
 and then take  $\Y$  such  that it has a non-trivial 
 intersection with the fixed finite subset $\X$.)    
\end{lem}

\begin{proof}
First we will  see  that $\Lchas$ is finite.
\begin{align*}
&\Lchas = \sup_{i  \in \Znu}
\sum_{\substack{\Y;\, \Y\cap\X\ne\emptyset \\ \chas(\X)\ne 0}}
\; \; \sum_{\substack{\X;\, \X \ni  i\\ \chas(\X)\ne 0}}\
\| \chas(\Y) \| \cdot \|\chas(\X) \|   \nonum\\
&\le \nonum\\
& \!\!\! \sup_{\X; \, \chas(\X)\ne 0} 
\Bigl| \bigl\{\Y; \, \Y\cap\X \ne\emptyset,\;
 \chas(\Y)\ne 0    \bigr\}  \Bigr| \cdot
\sup _{i  \in \Znu}
\Bigl| \bigl\{\X; \, \X\ni i,\;
\chas(\X)\ne 0  \bigr\} \Bigr| \cdot 
\| \chas \|_{\infty} \cdot  \| \chas \|_{\infty},
\end{align*}
 where $\bigl| \{\  \} \bigr|$ denotes 
 the cardinality of the set.
 By using  \eqref{eq:range} \eqref{eq:chasnorm} 
we further estimate  $\Lchas$ as follows. 
\begin{align}
\label{eq:Lchas-est-1st}
&\; \Lchas \nonum\\
&\le\!\! 
\sup_{\X;\, \diam(\X) \le r}\!
\Bigl| \bigl\{\Y; \, \Y\cap\X \ne\emptyset, \;
 \chas(\Y)\ne 0   \bigr\}  \Bigr| \cdot
\sup _{i  \in \Znu}  
\Bigl| \bigl\{
\X; \, \X\ni  i, \; \diam(\X) \le r \bigr\} \Bigr| 
\cdot  \| \chas \|_{\infty}^{\;2}\nonum\\
&\le \!\!\sup_{\X;\, \diam(\X) \le r}
\sum _{j \in  \X}
\Bigl| \bigl\{\Y; \, \Y\ni j,\; \diam(\Y) \le r   \bigr\}
\Bigr|\cdot \sup _{i  \in \Znu}  
\Bigl| \bigl\{\X; \, \X\ni  i, \; \diam(\X) \le r \bigr\} \Bigr| 
\cdot \| \chas \|_{\infty}^{\; 2}   \nonum\\
&\le \!\!\!\! \sup_{\X;\,  \diam(\X) \le r}
\!\!\!\!\!\! |\X| \cdot \max _{j \in  \X}
\Bigl| \bigl\{\Y; \, \Y\ni j, \,  \diam(\Y) \le r   \bigr\} \Bigr|
\cdot 
\sup _{i  \in \Znu}  
\Bigl| \bigl\{\X; \, \X\ni  i, \;
 \diam(\X) \le r \bigr\} \Bigr| 
\cdot
\| \chas \|_{\infty}^{\;2}   \nonum\\
&\le  |\Cubr| \cdot \sup _{j \in  \Znu}
\Bigl| \bigl\{\Y; \, \Y\ni j, \;\diam(\Y) \le r   \bigr\} \Bigr|
\cdot 
\sup _{i  \in \Znu}  
\Bigl| \bigl\{\X; \, \X\ni  i, \;
 \diam(\X) \le r \bigr\} \Bigr| 
\cdot
\| \chas \|_{\infty}^{\;2}   \nonum\\
&= \left(  \sup _{i  \in \Znu}  
\Bigl| \bigl\{
\X; \, \X\ni  i, \;
 \diam(\X) \le r \bigr\} \Bigr| 
\right)^{2}\cdot 
|\Cubr| \cdot \| \chas \|_{\infty}^{\;2} \nonum\\
&\le \Bigl| \bigl\{\X; \;
 |\X|\le |\Cubtwor| \bigr\} \Bigr|^{2} 
\cdot |\Cubr| \cdot  \| \chas \|_{\infty}^{\;2}  =
\left(2^{|\Cubtwor|}\right)^{2} \cdot |\Cubr|\cdot   
 \| \chas \|_{\infty}^{\;2}
=   4^{|\Cubtwor|}\cdot |\Cubr|  
\cdot  \| \chas \|_{\infty}^{\;2}\nonum\\
&\le 4^{|\Cubtwor|}\cdot |\Cubr|  
\cdot  (2\| \cha \|_{\infty})^{\;2}
= 4^{(2r+1)^{\nu}}\cdot (r+1)^{\nu}  
\cdot 4 \|  \cha \|_{\infty}^{\;2}
= 4^{(2r+1)^{\nu}+1}\cdot (r+1)^{\nu}  
\cdot  \|  \cha \|_{\infty}^{\;2}\nonum\\
&<\infty.
\end{align}
 
Let both $n\in\NN$ and $\I\Subset\Znu$ be fixed.
  For every  $A\in \FlI$
 by using \eqref{eq:derhcha-pair} in  Lemma \eqref{lem:pair}  repeatedly 
  we have  
\begin{align}
\label{eq:derhnA-prepare}
 &\Derhchan(A)\!\!\ \nonum\\
&=\sum_{\X_{2n-1}\cap \X_{2n} \ne\emptyset}
\cdots  \sum_{\X_3\cap \X_4 \ne\emptyset }
 \sum_{\X_1\cap \X_2 \ne\emptyset } \nonum\\
&\ \Bigl[\chas(\X_{2n})\chas(\X_{2n-1})
, \;\cdots {\Bigl[} \chas(\X_4)\chas(\X_3),\;  
\Bigl[\chas(\X_2)\chas(\X_1), \;  A\Bigr] 
{\Bigr]} \cdots \Bigl]\nonum\\
&= 
\sum_{ \bigl\{ \left(\X_1, \X_2, \cdots,\X_{2n-1}, \X_{2n}\right)\,|\,
(\X_{2i-1}\cup \X_{2i})  \cap \V_{2i-2}\ne \emptyset
\ {\text{and}}\ \X_{2i-1}\cap\X_{2i}
\ne \emptyset\ {\text{for every}} \ i \in \{1,2,\cdots, n\} \bigr\}}
\nonum\\
&\ \Bigl[\chas(\X_{2n})\chas(\X_{2n-1})
, \;\cdots {\Bigl[} \chas(\X_4)\chas(\X_3),\;  
\Bigl[\chas(\X_2)\chas(\X_1), \;  A\Bigr] 
{\Bigr]} \cdots \Bigl],
\end{align}
where  the 
second summation formula 
 is due to  the  locality
 and  the  notation   \eqref{eq:Vj} is used.
Similarly,  for each $\Lam \Subset \Znu$,
 by using \eqref{eq:derhchalam-pair} in  Lemma \eqref{lem:pair}  
 we obtain   for every  $A\in \FlI$
\begin{align}
\label{eq:derhlamnA-prepare}
 &\Derhchalamn(A)= \nonum\\
& \!\!\!\!\!\!\!\!\!\!\!\!
\sum_{ \bigl\{ \left(\X_1, \X_2, \cdots,\X_{2n-1}, \X_{2n}\right)\,|\,
(\X_{2i-1}\cup \X_{2i})  \cap \V_{2i-2}\ne \emptyset,
\ \X_{2i-1}\cap\X_{2i}\ne \emptyset,\ \X_{2i-1}\cap\Lam
\ne \emptyset,\; \X_{2i}\cap\Lam
\ne \emptyset\ {\text{for}} \; \forall i \in \{1,2,\cdots, n\} \bigr\}}
\nonum\\
&\ \Bigl[\chas(\X_{2n})\chas(\X_{2n-1})
, \;\cdots {\Bigl[} \chas(\X_4)\chas(\X_3),\;  
\Bigl[\chas(\X_2)\chas(\X_1), \;  A\Bigr] 
{\Bigr]} \cdots \Bigl].
\end{align}

Let us estimate the norm of $\Derhchan(A)$   
by induction of  $n\in \NN$.
Take  any  $2(n-1)$ finite subsets 
 $\left(\X_1, \X_2, \cdots,\X_{2n-3}, \X_{2n-2}\right)$
that satisfy  the required conditions  
$(\X_{2i-1}\cup \X_{2i})  \cap \V_{2i-2}\ne \emptyset$
and  $\X_{2i-1}\cap\X_{2i}
\ne \emptyset$  for every $i \in \{1,2,\cdots, n-1\}$.
For  such  fixed
 $\left(\X_1, \X_2, \cdots,\X_{2n-3}, \X_{2n-2}\right)$  
let us  define the 
following multiple commutator{\footnote{The notation $B_{n-1}(A)$ 
 should be made precise by noting 
explicitly its  dependence on the set of $2(n-1)$ finite subsets, like  
 $B_{n-1}(A; \{\X_1, \X_2, \cdots,\X_{2n-3}, \X_{2n-2}\})$.  
However, for simplicity  we will employ this short-hand notation 
 as in the cited literature.}} 
\begin{align}
\label{eq:Bnm}
&B_{n-1}(A)\nonum\\
&:=
\Bigl[\chas(\X_{2n-2})\chas(\X_{2n-3}), \;\cdots {\Bigl[} \chas(\X_4)\chas(\X_3),\;  
\Bigl[\chas(\X_2)\chas(\X_1), \;  A\Bigr] 
{\Bigr]} \cdots \Bigl]\in \Fl(\V_{2n-2}), \nonum\\
&B_{0}(A):=A\in \FlI,
\end{align}
 where  we use the notation \eqref{eq:Vj} 
\begin{equation*}
\V_{2n-2}\equiv \X_{2n-2} \cup \X_{2n-3}\cup \cdots \cup \X_1 \cup \I.
\end{equation*}
Note that  $\Derhchanmone(A)$ is the sum of  
$B_{n-1}(A)$s over  all 
the   finite subsets 
 $\left(\X_1, \X_2, \cdots, \X_{2n-3}, \X_{2n-2}\right)$
 satisfying    the required conditions stated above.

 We shall estimate  each  term $\Derhcha\left(B_{n-1}(A)\right)$
 that  appears in the summation formula  \eqref{eq:derhnA-prepare}
 of $\Derhchan(A)$.
By  \eqref{eq:derhcha-pair} and  \eqref{eq:Bnm} 
we have  
\begin{align}
\label{eq:derh-Bnm}
 &\Derhcha\left(B_{n-1}(A)\right)\!\!\ \nonum\\
&=\sum_{\X_{2n-1}\cap \X_{2n} \ne\emptyset}
\Bigl[\chas(\X_{2n})\chas(\X_{2n-1}), \; B_{n-1}(A)\Bigr]\nonum\\
&= \!\!\!\!\!\!\!\!\!\!\!\!\!\!\!\!\!\!\!\!\!\!\!\!\!\!\!
\sum_{ \bigl\{ \left(\X_{2n-1}, \X_{2n}\right)\,|\,
(\X_{2n-1}\cup \X_{2n})  \cap \V_{2n-2}\ne \emptyset
\ {\text{and}}\ \X_{2n-1}\cap\X_{2n}\ne \emptyset  \bigr\}}
\!\!\!\!\!\!\!\!\!\!\!\!\!\!\!\!\!\!\!\!\!\!\!\! 
\Bigl[\chas(\X_{2n})\chas(\X_{2n-1}), \; B_{n-1}(A)\Bigr],
\end{align}
where 
$\chas(\X_{2n})\chas(\X_{2n-1})\in \coree$  and 
the locality are  noted   in the second equality.
By using  \eqref{eq:Bnm} and \eqref{eq:derh-Bnm}  
we proceed the estimate as 
\begin{align}
\label{eq:derh-Bnm-norm}
 &\bigl\| \Derhcha\left(B_{n-1}(A)\right)\bigr\| \!\!\! \nonum\\
&\le 2  \| B_{n-1}(A) \| \cdot \!\!\!\!\!\!\!\!\!\!\!\!\!\!\!\!\!\!\!\!\!\!\!\!\!\!\!\!\!\!\!\!\!
\sum_{ \bigl\{ \left(\X_{2n-1}, \X_{2n}\right)\,|\,
(\X_{2n-1}\cup \X_{2n})  \cap \V_{2n-2}\ne \emptyset
\ {\text{and}}\ \X_{2n-1}\cap\X_{2n}\ne \emptyset  \bigr\}}
\!\!\!\!\!\!\!\!\!\!\!\!\!\!\!\!\!\!\!\!\!\!
\| \chas(\X_{2n}) \|
\cdot \| \chas(\X_{2n-1}) \| \nonum\\
%%%
&\le 2  \| B_{n-1}(A) \| \cdot
 \Bigl\{
\sum_{\X_{2n};\, \X_{2n}\cap\X_{2n-1}\ne\emptyset}
\;\;
\sum_{\X_{2n-1};\, \X_{2n-1}\cap\V_{2n-2}\ne\emptyset}
\| \chas(\X_{2n}) \|
\cdot \| \chas(\X_{2n-1}) \| \nonum\\ 
&\ \ \quad\quad +
\sum_{\X_{2n-1};\, \X_{2n-1}\cap\X_{2n}\ne\emptyset}
\;\;
\sum_{\X_{2n};\, \X_{2n}\cap\V_{2n-2}\ne\emptyset}
\| \chas(\X_{2n}) \|
\cdot \| \chas(\X_{2n-1}) \| \Bigr\}  
\nonum\\
&\le 2  \| B_{n-1}(A) \|
\cdot \bigl|\V_{2n-2}\bigr|\cdot 
 \Bigl\{\max_{i  \in \V_{2n-2}} \!
\sum_{\X_{2n};\, \X_{2n}\cap\X_{2n-1}\ne\emptyset}
\;\;
\sum_{\X_{2n-1} \ni  i}
\| \chas(\X_{2n}) \|
\cdot \| \chas(\X_{2n-1}) \| \nonum\\ 
&\quad\quad\quad\quad\quad\quad\quad\quad\quad+ 
\max_{i  \in \V_{2n-2}}\!
\sum_{\X_{2n-1};\, \X_{2n-1}\cap\X_{2n}\ne\emptyset}
\;
\sum_{\X_{2n} \ni  i}
\| \chas(\X_{2n}) \|
\cdot \| \chas(\X_{2n-1}) \| \Bigr\}  
\nonum\\ 
&\le 2  \| B_{n-1}(A) \|
\cdot \bigl|\V_{2n-2}\bigr|\cdot 
 2\Bigl\{\sup_{i  \in \Znu}
\sum_{\X_{2n};\, \X_{2n}\cap\X_{2n-1}\ne\emptyset}
\;
\sum_{\X_{2n-1} \ni  i}
\| \chas(\X_{2n}) \|
\cdot \| \chas(\X_{2n-1}) \| \Bigr\} \nonum\\ 
\nonum\\ 
&= 2 \| B_{n-1}(A) \| 
\cdot \bigl|\V_{2n-2} \bigr| \cdot  2 \cdot \Lchas  
= 4 \Lchas  \bigl|\V_{2n-2} \bigr| \cdot 
\| B_{n-1}(A) \| \nonum\\ 
&\le 4 \Lchas \Bigl( |\I|+2(n-1) \cdot (r+1)^{\nu}\Bigr)
\cdot \| B_{n-1}(A) \|, 
\end{align}
where we recall the notation  
$\Lchas$  defined in \eqref{eq:Lchas}
 and the estimate of 
$\bigl|\V_{2n-2} \bigr|$ given in 
\eqref{eq:Vjvol}.
 Note  that $B_{n-1}(A)$ of  \eqref{eq:Bnm}
 denotes  an arbitrary  term in the summation formula
 of $\Derhchanmone(A)$.   
Hence  by iteration we get 
\begin{align}
\label{eq:derhchanA-est}
 \bigl\| \Derhchan(A)\bigr\| &\le  (4\Lchas)^{n} \| B_{0}(A) \| 
\prod_{k=1}^{n}
\Bigl( |\I|+2(k-1) \cdot (r+1)^{\nu}\Bigr)\nonum\\
&\le  (4\Lchas)^{n} \| A \| 
\Bigl( |\I|+2(n-1) \cdot (r+1)^{\nu}\Bigr)^{n}
\end{align}
By noting the inequality  $\frac{x^{n}}{n!}\le e^{x}$  for 
$x\ge0$ we have the estimate  
\begin{align}
\label{eq:exptaylor}
&\frac{1}{n!} \Bigl( 4^{n} \Lchasn  
\bigl\{|\I|+2(n-1)  \cdot (r+1)^{\nu}\bigr\}^{n}  \Bigr)
=\frac{1}{n!}\Bigl(4\bigl\{|\I|+ 2(n-1)
\cdot (r+1)^{\nu}\bigr\} \Lchas \Bigr)^{n}\nonum\\ 
&\le\exp\Bigr(4\bigl\{|\I|+  2(n-1)  \cdot (r+1)^{\nu}\bigr\} \Lchas
\Bigl)=\exp\Bigr(4  |\I|  \Lchas \Bigr)
\exp\Bigr(8 (n-1) (r+1)^{\nu} \Lchas  \Bigl)\nonum\\
&=\exp\Bigr(4  \bigl\{|\I|-2(r+1)^{\nu}\bigr\}  \Lchas \Bigr) 
\exp\Bigr(n  \bigl\{8 (r+1)^{\nu} \Lchas \bigr\} \Bigl).
\end{align}
From the above estimates \eqref{eq:derhchanA-est}
 \eqref{eq:exptaylor} we obtain the estimate \eqref{eq:derhnA-est}
by setting  the real numbers $\kchas$ and $\Mchas$ as in 
 \eqref{eq:Lchas}.

In the same  way as we have shown the estimate  \eqref{eq:derhchanA-est},
 from the formula \eqref{eq:derhlamnA-prepare}  
we obtain  the following  norm estimate of $\Derhchalamn(A)$ 
independently of $\Lam\Subset \Znu$: 
 \begin{align}
\label{eq:derhchalamnA-prod}
 \bigl\| \Derhchalamn(A)\bigr\|
&\le  (4\Lchas)^{n} \| A \| 
\prod_{k=1}^{n}
\Bigl( |\I|+2(k-1) \cdot (r+1)^{\nu}\Bigr)\nonum\\
&\le   (4\Lchas)^{n} \| A \| 
\Bigl( |\I|+2(n-1) \cdot (r+1)^{\nu}\Bigr)^{n}.
\end{align}
This leads to the desired estimate  \eqref{eq:derhlamnA-est}
 by repeating  a similar argument given   above. 
\end{proof}

We  now give   a global time evolution.
\begin{prop}
\label{prop:alchatexist}
Let  $\cha\in \Chaspn$,
 i.e.  any  uniformly  bounded
nilpotent  finite-range  assignment of local fermion charges
 on the  fermion lattice system  of Definition {\rm{\eqref{defn:cha}}}.
  Let  $\Derhcha$ denote the  derivation  
defined  on $\core$  as  in  Definition {\rm{\eqref{defn:derhcha}}}.
Then there exists a strongly continuous one parameter
group of $\ast$-automorphisms  
$\{\alchat\,;\;t\in \R\}$
 on $\Fl$ given  by 
\begin{equation}
\label{eq:altglobal}
\alchat(F):= \Lamlim  \altlam(F)
{\text {\ \ in norm for each  \ }} F  \in \Fl
 {\text {\ \ and  \ }} t  \in \R,
 \end{equation}
 where $\altlam\equiv e^{it \Derhchalam}$  
as defined in \eqref{eq:altlam}. For each $F\in\Fl$ 
 the  convergence of \eqref{eq:altglobal}
is uniform  with respect to the parameter $t\in\R$ on  
any compact subset of $\R$.
Furthermore  $\Derhcha$ is a pre-generator of 
 $\{\alchat\,;\;t\in \R\}$ given above.
Namely, the norm closure of $\Derhcha$ 
 is exactly equal to  the generator of $\{\alchat\,;\;t\in \R\}$.
\end{prop}

\begin{proof}
By  \eqref{eq:vlimderhlam} 
$\Lamlim\Derhchalam(A)=\Derhcha(A)$ holds for every  $A\in \core$.
Let  $A\in \core$ and let $\I$ be a finite subset of $\Znu$
 such that $A\in \FlI$.
Let 
\begin{align}
\label{eq:tcirc}
t_{\circ}:= \frac{1}{ \Mchas},
\end{align}
 where  $\Mchas$  denotes the positive number given in 
 \eqref{eq:Lchas} in Lemma \eqref{lem:multiple-est}.
By   \eqref{eq:derhnA-est}
 the norm of $ \sum_{n=0}^{\infty} \frac{t^{n}}{n!} 
\Derhchan(A)$ is estimated from the above by 
$C\sum_{n=0}^{\infty } \left| \frac{t}{t_{0}} \right|^{n}$, where 
$C$ is some  constant determined solely by $A\in \core$ and $\cha\in \Chaspn$.
So it is convergent for $|t|< t_{\circ}$. 
Hence  any element of  $\core$ is  analytic for  $\Derhcha$.
 Using  the  argument given in  
 \cite[Proposition 6.2.3. Theorem 6.2.4]{BR}  
  we obtain  the statement.
\end{proof}

We now  redefine  `$\Derhcha$' as the generator 
of the  strongly continuous one parameter
group of $\ast$-automorphisms  $\{\alchat\,;\;t\in \R\}$ of $\Fl$.
\begin{defn}
\label{defn:derhchaclo}
For  each  $\cha\in \Chaspn$,  
the infinitesimal generator 
of the  strongly continuous one parameter
group of $\ast$-automorphisms  
$\{\alchat\,;\;t\in \R\}$
 on $\Fl$  given in Proposition 
 {\rm{\eqref{prop:alchatexist}}} is denoted as 
   $\Derhcha$.  
(From now on   $\Derhcha$   
denotes  the norm closure of the pre-generator 
defined  on $\core$  given in 
 Definition {\rm{\eqref{defn:derhcha}}}.)
\end{defn}

\subsubsection{Basic properties of superderivations on 
the fermion lattice  system}
\label{subsubsec:properties}
In the proof of   Lemma \eqref{lem:pair}  we do  not 
actually  use the nilpotent condition \eqref{eq:nil-delcha} of  
$\delcha$ and  the hermite property of $\delchas$, either. 
Only the identity $\Derhcha={\delchas}^{\!\!\!\!2}$ is  used there.
By noting  this fact  we  can  
 invent  a  criterion for the nilpotent condition   
  of $\delcha$  in terms  of $\cha$.
\begin{prop}
\label{prop:delcha-nil}
Suppose that  $\cha$ is a finite-range  assignment of local fermion charges
 on the  fermion lattice system  of Definition {\rm{\eqref{defn:cha}}}.
 Let $\delcha$ denote the superderivation  associated with   $\cha$.
Then for each $A\in \core$, 
\begin{align}
\label{eq:deldelA-pair}
\delcha\cdot\delcha(A)
=\frac{1}{2} \sum_{\substack{\X \Subset \Znu, \, \Y \Subset \Znu
\\  \X \cap \Y \ne\emptyset}}
\Bigl[  \bigl\{\cha(\X),\;   \cha(\Y)\bigr\}, \; A \Bigl]
=\sum_{\substack{\X \Subset \Znu, \, \Y \Subset \Znu
\\  \X \cap \Y \ne\emptyset}}\Bigl[ \cha(\X) \cha(\Y),\;  A \Bigr].
\end{align}
If $\cha(\X)$ and $\cha(\Y)$ anti-commute  for any (non-disjoint)  
pair of $\X \Subset\Znu$ and $\Y \Subset\Znu$,
 then $\delcha$  is nilpotent.
\end{prop}
\begin{proof}
We  can derive the  formula \eqref{eq:deldelA-pair}
in the same way as   the   formula \eqref{eq:derhcha-pair}
 in  Lemma \eqref{lem:pair}  by taking $\cha$ in the place of $\chas$.
If $\bigl\{\cha(\X),\;   \cha(\Y)\bigr\}=0$
for any  $\X \Subset\Znu$ and $\Y\Subset\Znu$, 
 then  by the formula   \eqref{eq:deldelA-pair}
 $\delcha\cdot\delcha=\zeromap$ on $\core$.
\end{proof}

\begin{rem} 
\label{rem:spinOK}
  Proposition \eqref{prop:delcha-nil}
   is valid  when the spin  of fermions exists.
It is valid for other lattices as well.
\end{rem}

\begin{rem} 
\label{rem:HAGEN}
 A condition of superderivations 
on the fermion  lattice system
for being  nilpotent   has  been given  under 
the periodic boundary condition imposed in  \cite{HAGEN13}.
 Proposition \eqref{prop:delcha-nil}
 does not  require  any  specific   boundary condition on finite systems.
 The superderivation $\delcha$  
 and the  finite-range  assignment of local fermion charges
$\cha$   are linked together 
 by the  relation \eqref{eq:DEFdelcha}. 
\end{rem}

We will show  norm-closability   of $\delcha$, as 
 it  is  a fundamental property required 
 for  reasonable  supersymmetric $\cstar$-dynamics.
For some technical reason, we assume that  supersymmetry is 
unbroken, namely $\cha$ is in  $\Chaspnun$.
\begin{prop}
\label{prop:allclosable-unbroken}
Let $\cha\in \Chaspnun$. 
Then the associated nilpotent superderivations 
$\delcha$ and $\delchaast$ defined on $\core$   
are  norm-closable. Similarly the  hermite  superderivations 
  $\delchasf$ and $\delchass$ defined on $\core$ are norm-closable. 
\end{prop}
\begin{proof}
By the assumption there exists a supersymmetric state  $\vp$   
 with respect to $\delcha$.
By Proposition \eqref{prop:susy-con},  
 it is  invariant under  $\delchaast$ on  
$\core$ as well. Similarly by Proposition \eqref{prop:inv-symdel},  
  it is invariant under $\delchasf$ and  $\delchass$ on $\core$. 
We note that the total system $\Fl$ is 
 simple, as it is the  CAR algebra. 
Hence  by applying  Proposition \eqref{prop:faithclo}
 to the superderivations 
  $\delcha$, $\delchaast$, $\delchasf$ and 
$\delchass$  it is concluded that they   are all  norm-closable.
\end{proof}

\subsubsection{Global time evolution commutes with superderivations}
\label{subsubsec:kakan}
 Based on the commutativity 
between supersymmetric Hamiltonian and supercharges
  as  in  \eqref{eq:HQkakan} in Sect.\ref{sec:overview}, it is usually 
 said that  supersymmetry is   a symmetry.{\footnote{Compare this extended usage  with   the notion of symmetries by Wigner.   
 We refer to e.g.  Sect.8.1 of \cite{STRbook}.}}
Since  it  is a  direct consequence  of  the  supersymmetry algebra,  
 we expect a similar commutativity relation 
 for supersymmetric  fermion lattice models. 
In fact,  
  we will show below  that the global time evolution 
$\{\alchat\,;\;t\in \R\}$ commutes with  the  superderivations
 $\delcha$ and  $\delchaast$.
It requires   non-trivial  analysis.

\begin{prop}
\label{prop:koukan-lat}
Let  $\cha\in \Chaspn$.
  Let $\delcha$ denote the superderivation   
defined on $\core$  associated with  $\cha$.
Let  $\delchas$ denote each of the  hermite superderivations  
 $\delchasf$  and   $\delchass$  defined on $\core$  in \eqref{eq:delchas-lat}. Let $\{\alchat\,;\;t\in \R\}$ denote the strongly continuous one parameter 
 group of $\ast$-automorphisms  given in  
Proposition  {\rm{\eqref{prop:alchatexist}}}.  
If $\delchas$ is norm-closable, then 
\begin{equation}
\label{eq:koukan-fer-sym}
\alchatcore \subset \domdelchascl, \quad 
\delchascl \cdot \alchat=\alchat \cdot \delchas 
{\text{\ \ on\ }} \core \ \ {\text{for  each}}\  t\in \R,
 \end{equation}
where  $\delchascl$  denotes  the norm-closure of  $\delchas$
and  $\domdelchascl$ denotes the  domain of $\delchascl$.
If the nilpotent superderivation   $\delcha$ is norm-closable, then 
\begin{equation}
\label{eq:koukan-fer}
\alchatcore \subset \domdelchacl,\quad
\delchacl \cdot \alchat=\alchat \cdot \delcha 
{\text{\ \ on\ }} \core \ \ {\text{for  each}}\  t\in \R,
 \end{equation}
and 
\begin{equation}
\label{eq:koukan-fer-ast}
\alchatcore \subset \domdelchaastcl,\quad
\delchaastcl \cdot \alchat=\alchat \cdot \delchaast 
{\text{\ \ on\ }} \core \ \ {\text{for  each}}\  t\in \R.
 \end{equation}
\end{prop}

\begin{proof}
Let   $r$ denote the range of $\cha$.
Let $A$ denote an arbitrary element of $\core$ which will be fixed 
in what follows.
Let  $\I$ denote the smallest  finite subset such that 
 $A\in \FlI$.
Take any finite subset $\Lam$ that includes $\I$.
(Later we will let $\mugen$.)
From  \eqref{eq:derhchalamn} and $\I\cup \Lamr=\Lamr$   
it follows  that
 $\Derhchalamn(A)\in \Fl(\Lamr)$  for every   $n\in\NN$, and 
from  \eqref{eq:altlam-affil}
   $\altlam(A)\in \Fl(\Lamr)$  for every  $t\in \R$.
By  noting \eqref{eq:delchas-lat} \eqref{eq:derhchalamn} 
  we have for each  $n\in \NN$    
\begin{align}
\label{eq:delchas-derhchalamn-koukan}
&\delchas \bigl(\Derhchalamn(A)\bigr) \nonum\\
&=\sum_{\X} \;\Bigl[\chas(\X), \;\Derhchalamn(A) \Bigr]_{\gamma} 
=\sum_{\X \cap \Lamr \ne \emptyset} 
\;\Bigl[\chas(\X), \;\Derhchalamn(A) \Bigr]_{\gamma}\nonum\\
&=\sum_{\X \cap  \Lamr \ne \emptyset} \;
\sum_{\X_1\cap \Lam \ne\emptyset, \cdots ,\X_{2n}\cap \Lam \ne\emptyset}
\Bigl[\chas(\X), \;
\Bigl[\chas(\X_{2n}), \;\cdots   \Bigl[\chas(\X_1), \;  A\Bigr]_{\gamma} 
 \cdots \Bigl]_{\gamma} \Bigl]_{\gamma}
\nonum\\
&=\sum_{\X \cap  \Lam \ne \emptyset}\; 
\sum_{\X_1\cap \Lam \ne\emptyset,\; 
 \cdots ,\X_{2n}\cap \Lam \ne\emptyset}
\Bigl[\chas(\X),\;
\Bigl[\chas(\X_{2n}), \;\cdots  \Bigl[\chas(\X_1), \;  A\Bigr]_{\gamma} 
 \cdots \Bigl]_{\gamma}\Bigl]_{\gamma}\nonum \\
&\ \ +
\sum_{\substack{\X \cap  \Lamr \ne \emptyset
\\     \X \cap  \Lam = \emptyset} }\;
\sum_{\X_1\cap \Lam \ne\emptyset,\; 
 \cdots ,\X_{2n}\cap \Lam \ne\emptyset}
\Bigl[\chas(\X), \;
\Bigl[\chas(\X_{2n}), \;\cdots \Bigl[\chas(\X_1), \;  A\Bigr]_{\gamma} 
 \cdots \Bigl]_{\gamma}\Bigl]_{\gamma}
\nonum\\
&=\!\!\!\sum_{\X \cap  \Lam \ne \emptyset}\, 
\sum_{\X_2\cap \Lam \ne\emptyset, \cdots ,\X_{2n}\cap \Lam \ne\emptyset}
\!\!\!\!\!\!\!\!\!\!\!\!
\Bigl[\chas(\X), \;
\Bigl[\chas(\X_{2n}), \;\cdots {\Bigl[} \chas(\X_2),\;  
\!\!\sum_{\X_1\cap \Lam \ne\emptyset}\!
\Bigl[\chas(\X_1), \;  A\Bigr]_{\gamma} 
{\Bigr]}_{\gamma} \cdots \Bigl]_{\gamma}\Bigl]_{\gamma}\nonum \\
&\, +\!\!
\sum_{\substack{\X \cap  \Lamr \ne \emptyset
\\     \X \cap  \Lam = \emptyset}}\,
\sum_{\X_1\cap \Lam \ne\emptyset,\; 
\X_2\cap \Lam \ne\emptyset, \cdots ,\X_{2n}\cap \Lam \ne\emptyset}
\!\!\!\!\!\!\!\!
\!\!\!\!\!\!\!\!\!\!\!\!\!\!
\Bigl[\chas(\X),\;
\Bigl[\chas(\X_{2n}), \;\cdots {\Bigl[} \chas(\X_2),\;\Bigl[\chas(\X_1), \;  A\Bigr]_{\gamma} 
{\Bigr]}_{\gamma} \cdots \Bigl]_{\gamma}\Bigl]_{\gamma} 
\nonum\\
&=\!\!\!
\sum_{\X_2\cap \Lam \ne\emptyset, \cdots ,\X_{2n}\cap \Lam \ne\emptyset, 
\,\X \cap  \Lam \ne \emptyset}
\!\!\!\!\!\!\!\!
 \Bigl[\chas(\X), \;
\Bigl[\chas(\X_{2n}), \;\cdots {\Bigl[} \chas(\X_2),\;  
\delchas(A) {\Bigr]}_{\gamma} \cdots {\Bigr]}_{\gamma}  \Bigl]_{\gamma}
\nonum\\
 &\quad\quad \quad +
\sum_{\X \cap  \Lamr \ne \emptyset,\, 
\X \subset   \Lamc}\;
\Bigl[\chas(\X), \;
\Derhchalamn(A) \Bigr]_{\gamma} 
\nonum\\
&= \Derhchalamn\bigl(\delchas(A)\bigr)+\bnlam(A),
 \end{align}
where  
$\Lamc$ denotes the  complement of $\Lam$ in $\Znu$
 and  we have defined 
\begin{align}
\label{eq:bnlam}
\bnlam(A):=
\sum_{\X \cap  \Lamr \ne \emptyset,\, 
\X \subset   \Lamc}\;
\Bigl[\chas(\X), \;
\Derhchalamn(A) \Bigr]_{\gamma}.
\end{align}

We will show that  there exist  a constant  $\kpchas$ 
depending on $A\in \core$ and  $\chas$,   
and a constant $\Mchas$ depending  on  $\chas$  satisfying  that   
\begin{align}
\label{eq:bnlam-uekara}
 \frac{1}{n!}\bigl\| \bnlam(A)\bigr\| 
\le  \kpchas \cdot \Mchasn. 
\end{align}
By   the definition  of $\bnlam$ in \eqref{eq:bnlam}
 and  the summation formula 
 of $\Derhchalamn(A)$  in  \eqref{eq:derhlamnA-prepare}  
  we see  that 
\begin{align}
\label{eq:bnlam-tenkai}
&\bnlam(A)\nonum\\
&=\sum_{\X \cap  \Lamr \ne \emptyset,\, \X \cap  \Lam = \emptyset }\nonum\\
& \sum_{\bigl\{ \left(\X_1, \X_2, \cdots ,\X_{2n}\right)\,|\,
(\X_{2i-1}\cup \X_{2i})  \cap \V_{2i-2}\ne \emptyset,
\;\X_{2i-1}\cap\X_{2i}
\ne \emptyset,\; 
\X_{2i-1}\cap  \Lam \ne \emptyset,\; 
\X_{2i}\cap  \Lam \ne \emptyset \; 
{\text{for}}\ \forall i \in \{1,2,\cdots, n\} \bigr\}}  \nonum\\
&\quad \biggl[\chas(\X), \;\Bigl[\chas(\X_{2n})\chas(\X_{2n-1})
, \;\cdots {\Bigl[} \chas(\X_4)\chas(\X_3),\;  
\Bigl[\chas(\X_2)\chas(\X_1), \;  A\Bigr] 
{\Bigr]} \cdots \Bigl]\biggr]_{\gamma},
\end{align}
where  only  the last commutator involving  $\chas(\X)$ is 
the $\gamma$-graded commutator, while the others are  the  usual 
 commutator. 
  For each
 $\left(\X_1, \X_2, \cdots, \X_{2n-1}, \X_{2n}\right)$ that is relevant 
 for  the above sum, as in \eqref{eq:Bnm}, let 
\begin{align}
\label{eq:BnA}
B_{n}(A)&=
\Bigl[\chas(\X_{2n})\chas(\X_{2n-1}), \;\cdots {\Bigl[} \chas(\X_4)\chas(\X_3),\; \Bigl[\chas(\X_2)\chas(\X_1), \;  A\Bigr] 
{\Bigr]} \cdots \Bigl]\in \Fl(\V_{2n}),\nonum\\
\ B_{0}(A)&=A\in \FlI,
\end{align}
\begin{equation*}
\V_{2n}\equiv \X_{2n} \cup \X_{2n-1}\cup \cdots \cup \X_1 \cup \I.
\end{equation*}
Note that  the above $B_{n}(A)$ 
 depends on  the set of finite subsets
 $(\X_1, \X_2, \cdots ,\X_{2n-1}, \X_{2n})$.
Let us estimate 
$\sum_{\X \cap  \Lamr \ne \emptyset,\, \X \cap  \Lam = \emptyset}
\bigl[\chas(\X), \; B_{n}(A) \bigr]_{\gamma}$.
By \eqref{eq:BnA} together with 
the  $\gamma$-locality \eqref{eq:glocality} 
\begin{align*}
\sum_{\X \cap  \Lamr \ne \emptyset,\, \X \cap  \Lam = \emptyset }
\bigl[\chas(\X), \; B_{n}(A) \bigr]_{\gamma}  
= 
\sum_{\X \cap  \Lamr \ne \emptyset,\, 
\X \cap  \Lam = \emptyset,\,  \X\cap \V_{2n} \ne \emptyset}
\bigl[\chas(\X), \; B_{n}(A) \bigr]_{\gamma}.  
\end{align*}
Hence  we obtain 
\begin{align}
\label{eq:IrkyouBnA}
&\quad \left\| \sum_{\X \cap  \Lamr \ne \emptyset,\, \X \cap  \Lam = \emptyset }
\bigl[\chas(\X), \; B_{n}(A) \bigr]_{\gamma} \right\| \nonum\\
&\le 
\sum_{\X \cap  \Lamr \ne \emptyset,\, 
\X \cap  \Lam = \emptyset,\,  \X\cap \V_{2n} \ne \emptyset}
2 \| \chas(\X) \|
\cdot \| B_{n}(A) \| \nonum\\
&\le 
\sum_{\X \cap \V_{2n} \ne \emptyset}
2 \| \chas(\X) \|
\cdot \| B_{n}(A) \| \nonum\\
&\le \bigl|\V_{2n}\bigr| \cdot \max_{i\in  \V_{2n} }
\sum_{\X \ni  i}  2 \| \chas(\X) \| \cdot 
\| B_{n}(A) \|
\nonum\\
&\le 
 \bigl|\V_{2n}\bigr| \cdot 
2^{|\Cubtwor|} \cdot 2 \| \chas \|_{\infty}
\cdot \| B_{n}(A) \|
 \nonum\\
&\le 
\bigl(|\I|+2n  \cdot (r+1)^{\nu}\bigr)
\cdot 2^{(2r+1)^{\nu}} \cdot 2\| \chas \|_{\infty}\cdot 
\| B_{n}(A) \|.
\end{align}
In the same way as \eqref{eq:derh-Bnm-norm} for $\Derhcha$
we can show  a  similar estimate for 
$\Derhchalam$. Namely for each $k\in \NN$
\begin{align}
\label{eq:derhlam-Bnm-norm}
 \bigl\| \Derhchalam\left(B_{k}(A)\right)\bigr\| 
\le 4 \Lchas \Bigl( |\I|+2k \cdot (r+1)^{\nu}\Bigr)
\cdot \| B_{k}(A) \|. 
\end{align}
Now recall  the multi-summation  
formula  \eqref{eq:bnlam-tenkai} of  $\bnlam(A)$.
 By using \eqref{eq:derhlam-Bnm-norm} inductively   from $k=0$ to $k=n-1$
 and then applying  \eqref{eq:IrkyouBnA}
we get 
\begin{align}
\label{eq:bnlamest-product}
&\| \bnlam(A)\| \nonum \\ 
&\le \bigl(|\I|+2n  \cdot (r+1)^{\nu}\bigr)
\cdot 2^{(2r+1)^{\nu}} \cdot 2\| \chas \|_{\infty}\cdot 
 (4\Lchas)^{n} \| A \| 
\prod_{k=0}^{n-1}
\Bigl( |\I|+2k \cdot (r+1)^{\nu}\Bigr)\nonum\\
&= 2^{(2r+1)^{\nu}+1} \| \chas \|_{\infty} 
\cdot \| A \| \cdot (4\Lchas)^{n}  \cdot     
\bigl(|\I|+2n  \cdot (r+1)^{\nu}\bigr)
\cdot |\I| \cdot
\prod_{k=1}^{n-1}
\Bigl( |\I|+2k  \cdot (r+1)^{\nu}\Bigr)\nonum\\
&\le 
2^{(2r+1)^{\nu}+1} \| \chas \|_{\infty} 
\cdot \| A \| \cdot  |\I| \cdot  (4\Lchas)^{n}  
\Bigl( |\I|+2n \cdot (r+1)^{\nu}\Bigr)^{n}.
\end{align}
By  the inequality  $\frac{x^{n}}{n!}\le e^{x}$  for 
$x\ge0$ we have 
\begin{align}
\label{eq:exptaylor-bnlam}
&\frac{1}{n!} \Bigl( (4\Lchas)^{n}  
\bigl( |\I|+2n   \cdot (r+1)^{\nu}\bigr)^{n}  \Bigr)
=\frac{1}{n!}\Bigl(4\bigl\{|\I|+2n \cdot (r+1)^{\nu}\bigr\} 
\Lchas \Bigr)^{n}\nonum\\ 
&\le\exp\Bigr(4\bigl\{|\I|+  2n  \cdot (r+1)^{\nu}\bigr\} \Lchas
\Bigl)=\exp\Bigr(4  |\I|  \Lchas \Bigr)
\exp\Bigr(8 n (r+1)^{\nu} \Lchas  \Bigl)\nonum\\
&=\exp\Bigr(4  |\I|  \Lchas \Bigr) 
\exp\Bigr(n  \bigl\{8 (r+1)^{\nu} \Lchas \bigr\} \Bigl).
\end{align}

By setting  
\begin{align}
\label{eq:kpchas}
\kpchas:=
2^{(2r+1)^{\nu}+1} 
 \| \chas \|_{\infty}\cdot 
  \| A \|\cdot |\I| \cdot
\exp\Bigr(4  |\I|   \Lchas \Bigr),
\end{align}
and 
\begin{align}
\label{eq:Mchasforbn}
\Mchas:=
\exp\Bigr(8   (r+1)^{\nu} \Lchas  \Bigl),
\end{align}
 the desired estimate \eqref{eq:bnlam-uekara} holds for any $\Lam$.
(The above $\Mchas$ and $\Lchas$ are both  exactly  same as  those 
  in  \eqref{eq:Lchas}. Note that these   
 constants  are independent of $\Lam$.)

As noted before   $\altlam(A)$ and  $\Derhchalamn(A)$ are  in  
the finite subsystem $\Fl(\Lamr)$ on which 
 $\delchas$ is obviously  bounded.
 By using \eqref{eq:Fnlam}  \eqref{eq:delchas-derhchalamn-koukan} 
we have 
\begin{align}
\label{eq:koukan-gosa}
\delchas \bigl(\altlam(A)\bigr)&=
\delchas \left(\sum_{n=0}^{\infty}\frac{(it)^{n}}{n!}
\Derhchalamn(A)\right) 
= \sum_{n=0}^{\infty}\frac{(it)^{n}}{n!} \delchas \left(
\Derhchalamn(A)\right) 
\nonum\\
&=\sum_{n=0}^{\infty} \frac{(it)^{n}}{n!}
 \Bigl( \Derhchalamn\bigl(\delchas(A)\bigr)+  \bnlam(A)
\Bigr)\nonum\\
&=\sum_{n=0}^{\infty} \frac{(it)^{n}}{n!}
 \Derhchalamn\bigl(\delchas(A)\bigr)+ 
\sum_{n=0}^{\infty} \frac{(it)^{n}}{n!} \bnlam(A)\nonum\\
&=\altlam \bigl( \delchas(A)\bigr)+ 
\sum_{n=0}^{\infty} \frac{(it)^{n}}{n!} \bnlam(A).
\end{align}
As before let  $t_{\circ}:= \frac{1}{ \Mchas}>0$.
Due to  the estimate \eqref{eq:bnlam-uekara}, for an arbitrary  $\varepsilon>0$
there exists an  $N_{\circ}\in \NN$
independently of $\Lam\Subset\Znu$  such  that 
\begin{equation}
\label{eq:bnlam-kouhan-eps}
\sum_{n=N_{\circ}+1}^{\infty } 
\left\|   \frac{\bnlam(A)}{n!} t^{n} \right\|<\varepsilon
{\text{\ \ for \ }} |t|< t_\circ.
\end{equation}
Next we will show that  $\sum_{n=0}^{N_{\circ}}\frac{(it)^{n}}{n!} \bnlam(A)$ 
is negligible if we take   $\Lam$ sufficiently large.
By \eqref{eq:derhchalamn} we have $\Derhchalamn(A)\in  \Fl(\ItwoNcr)$
 for every    $n\in\{0,1,2,\cdots,N_{\circ}\}$.
By the definition  of $\bnlam(A)$  \eqref{eq:bnlam} 
and  the $\gamma$-locality, 
if  $\Lam\supset\ItwoNcr$, then $\bnlam(A)=0$
 for all  $n\in\{0,1,2, \cdots ,N_{\circ}\}$.  
Hence  
\begin{equation}
\label{eq:bnlam-zenhan-zero}
\Lamlim \sum_{n=0}^{N_{\circ}} 
 \frac{(it)^{n}}{n!} \bnlam(A) =0
{\text{\ \ for all\ }} t\in \R.
\end{equation}
By  \eqref{eq:bnlam-kouhan-eps}
and \eqref{eq:bnlam-zenhan-zero},  letting $\varepsilon\to 0$  
 we obtain  
\begin{equation}
\label{eq:vlimbnlam-zero}
\Lamlim  \sum_{n=0}^{\infty}
 \frac{(it)^{n}}{n!} \bnlam(A)  =0
{\text{\ \ for \ }} |t|< t_\circ.
\end{equation}
From  the above formulas  \eqref{eq:koukan-gosa} \eqref{eq:vlimbnlam-zero} 
combined with  \eqref{eq:altglobal} in Proposition 
 {\rm{\eqref{prop:alchatexist}}}
it follows that   
\begin{equation}
\label{eq:lamlimtcirc-koukan-s}
\Lamlim \delchas \bigl(\altlam(A)\bigr)=
  \alchat \bigl( \delchas(A)\bigr)
{\text{\ \ for \ }} |t|< t_\circ.
\end{equation}
If $\delchas$ is  norm-closable, then   
Eqs.\eqref{eq:lamlimtcirc-koukan-s} 
 \eqref{eq:altglobal} imply that  
  \begin{equation}
\label{eq:t-seigen-s}
 \alchat(A)\in \domdelchascl,\quad 
 \delchascl \bigl( \alchat(A)\bigr)=
  \alchat \bigl( \delchas(A)\bigr)
{\text{\ \ for   \ }}  |t|< t_\circ.
 \end{equation}

We shall  remove the  restriction of $t\in \R$ in  
the equality \eqref{eq:t-seigen-s}.
Let $A$ denote any element of  $\core$ as before. 
Take $s, t\in \R$  such that   $|s|< t_\circ$ and $|t|< t_\circ$.
By applying  the formula \eqref{eq:t-seigen-s}
directly to  $\alslam(A)\in \core$  
 for  any  fixed $\Lam\Subset\Znu$ we have   
\begin{equation}
\label{eq:remove-i}
 \alchat\bigl(\alslam(A)\bigr)\in \domdelchascl,\quad 
\delchascl \left( \alchat\bigl(\alslam(A)\bigr)\right)=
  \alchat \left( \delchas
\bigl(\alslam(A)\bigr)\right) {\text{\ \ for \ }} |t|< t_\circ.
 \end{equation}
As  $|s|< t_\circ$, by noting  \eqref{eq:lamlimtcirc-koukan-s},  
  taking $\mugen$ of the right hand side of \eqref{eq:remove-i} yields
\begin{align}
\label{eq:remove-add}
\Lamlim 
  \alchat \left( \delchas\bigl(\alslam(A)\bigr)\right) 
=  \alchat \left( \alchas \bigl( \delchas(A)\bigr) \right)
=\alchats \bigl( \delchas(A)\bigr).
 \end{align} 
Since  $\Lamlim \alchat\bigl(\alslam(A)\bigr)=  \alchats(A)$
 and   $\delchas$ is   norm-closable by the  assumption,
by  \eqref{eq:remove-i} and \eqref{eq:remove-add}
we have 
\begin{align}
\label{eq:stkoukan}
\alchats(A)\in \domdelchascl,\ 
 \delchascl \bigl(\alchats(A)\bigr)=
 \alchats \bigl( \delchas(A)\bigr) 
{\text{\ for  \ }} |s|,\,|t|< t_\circ.
 \end{align}
 Repeating   the above extension procedure 
from Eq.\eqref{eq:t-seigen-s} to Eq.\eqref{eq:stkoukan}  
 we obtain  the assertion  \eqref{eq:koukan-fer-sym}.

Next we will show  \eqref{eq:koukan-fer} and \eqref{eq:koukan-fer-ast}.
As  Eq.\eqref{eq:delchas-derhchalamn-koukan} holds  
 for both   $\chas=\chasf$ and $\chas=\chass$, by
 noting  \eqref{eq:delcha-gyaku} 
we  see that for every   $n\in \NN$   
\begin{align}
\label{eq:delcha-derhchalamn-koukan}
\delcha \bigl(\Derhchalamn(A)\bigr) 
&= \Derhchalamn\bigl(\delcha(A)\bigr)+\bnlambar(A), \nonum\\
\delchaast \bigl(
\Derhchalamn(A)\bigr) 
&= \Derhchalamn\bigl(\delchaast(A)\bigr)+\bnlamtil(A), 
 \end{align}
 where   we have defined 
\begin{align}
\label{eq:bnlambar}
\bnlambar(A)&:=
\sum_{\X \cap  \Lamr \ne \emptyset,\, 
\X \subset   \Lamc}\;
\Bigl[\cha(\X), \;
\Derhchalamn(A) \Bigr]_{\gamma},  \nonum\\
\bnlamtil(A)&:=
\sum_{\X \cap  \Lamr \ne \emptyset,\, 
\X \subset   \Lamc}\;
\Bigl[\chaast(\X), \;
\Derhchalamn(A) \Bigr]_{\gamma}.
\end{align}
Repeating  the argument  from  Eq.\eqref{eq:delchas-derhchalamn-koukan} 
to Eq.\eqref{eq:lamlimtcirc-koukan-s}  we have 
\begin{align}
\label{eq:lamlimtcirc-koukan-cha}
\Lamlim \delcha \bigl(\altlam(A)\bigr)=
\alchat \bigl( \delcha(A)\bigr) {\text{\ \ for \ }} |t|< t_\circ,
\nonum\\
\Lamlim \delchaast \bigl(\altlam(A)\bigr)=
  \alchat \bigl( \delchaast(A)\bigr) {\text{\ \ for \ }} |t|< t_\circ.
\end{align}
If  $\delcha$ is  norm-closable, then
 $\delchaast$ is also norm closable by
 Proposition \eqref{prop:ast-normclo}. Hence for this case,  
we obtain both    
\begin{align}
\label{eq:t-seigen-koukan-cha}
 \alchat(A)&\in \domdelchacl,\quad 
 \delchacl \bigl( \alchat(A)\bigr)=
  \alchat \bigl( \delcha(A)\bigr)
{\text{\ \ for  \ }}  |t|< t_\circ,
  \end{align}
and 
\begin{align}
\label{eq:t-seigen-koukan-chaast}
\alchat(A)&\in \domdelchaastcl,\quad 
 \delchaastcl  \bigl( \alchat(A)\bigr)=
  \alchat \bigl( \delchaast(A)\bigr)
{\text{\ \ for  \ }}  |t|< t_\circ.
\end{align}
We can remove the  restriction of $t\in \R$ in  
\eqref{eq:t-seigen-koukan-cha}  
\eqref{eq:t-seigen-koukan-chaast} 
  as before  and get   the desired formulas  
  \eqref{eq:koukan-fer} \eqref{eq:koukan-fer-ast}.      
\end{proof}

\begin{coro}
\label{coro:koukan-lat-unbroken}
Let  $\cha\in \Chaspnun$. 
Then all the commutativity relations 
between the global time evolution and the superderivations 
\eqref{eq:koukan-fer-sym} 
\eqref{eq:koukan-fer} \eqref{eq:koukan-fer-ast}
 in Proposition {\rm{\eqref{prop:koukan-lat}}} hold.
\end{coro}

\begin{proof}
The statement directly follows from the combination of 
Proposition \eqref{prop:allclosable-unbroken} and 
Proposition \eqref{prop:koukan-lat}.
\end{proof}

By  collecting   the results established in this section  
 we propose   a general class of  supersymmetric $\cstar$-dynamics 
 on the  fermion lattice system.
\begin{thm}
\label{thm:main}
Let  $\left(\Fl,\;  \{\FlI;\; \I\Subset \Znu \},\; \gamma\right)$
denote  the fermion  lattice system on $\Znu$.    
 Let $\core$ denote the  local algebra  of $\Fl$. 
 Take   any $\cha\in \Chaspn$  of   
 Definition \eqref{defn:cha}, namely 
any  uniformly  bounded  finite-range  assignment of  
local fermion charges on the  fermion lattice system
 whose associated superderivation $\delcha$  is nilpotent.
 Then there exists  a strongly continuous one parameter
group of $\ast$-automorphisms   $\{\alchat\,;\;t\in \R\}$
on the CAR algebra $\Fl$ whose generator $\Derhcha$ is equal to 
 the norm closure of the supersymmetric derivation in the form of 
$\delchaast\cdot\delcha+ \delcha\cdot\delchaast$ on $\core$. 
Furthermore if $\cha\in \Chaspnun$ 
of Definition  {\rm{\eqref{defn:chaunbroken}}},
then  the nilpotent superderivations $\delcha$ and  $\delchaast$,
and  the hermite superderivations $\delchasf$ and $\delchass$  
all commute  with  the time evolution $\{\alchat\,;\;t\in \R\}$ on $\core$.
\end{thm}

\begin{rem} 
\label{rem:KMS}
Consider  $\cha\in \Chaspnun$ 
of Definition \eqref{defn:chaunbroken}.
 Then  any  supersymmetric state $\vp$ 
 with respect to $\delcha$  (in the sense of  Definition \eqref{defn:SUSYSTATE})
 is a ground state for $\{\alchat\,;\;t\in \R\}$, because 
$\core$ is a core for the generator $\Derhcha$ 
 of  $\{\alchat\,;\;t\in \R\}$  by   Definition \eqref{defn:derhchaclo}, 
 and the required condition  to be a ground state 
(Definition 5.3.18 of \cite{BR})  is satisfied as follows: 
\begin{align}
\label{eq:ground}
\vp\left( A^{\ast} \Derhcha(A) \right)&= 
\vp\left( A^{\ast} 
\delchaast (\delcha(A)) \right)
+\vp\left( A^{\ast} 
\delcha (\delchaast(A)) \right) \nonum\\
&=\vp\left( \delcha(A)^{\ast} \delcha(A) \right)
+\vp\left(\delchaast(A)^{\ast} 
\delchaast(A) \right)\ge 0, \quad   \forall A\in\core.
\end{align}
See   Proposition 2.2 of \cite{BU} for the detail.
\end{rem}

\section{Examples} 
\label{sec:EX}
In this section we  present some  
  supersymmetric fermion lattice models in the $\cstar$-algebraic 
formulation  stated in  Sect.\ref{sec:SUSYfer}.
For  the  first  example  we will reproduce 
 the  Nicolai's model   \eqref{eq:Q-intro} \eqref{eq:H-intro}
in  Sect.\ref{sec:overview} in the   $\cstar$-algebraic format.
Define  a map from  $\{\I; \; \I\Subset  \Z\}$
into $\coreo$  by   
\begin{align}
\label{eq:chanic}
\chanicisan&:= \atwoip \atwoicr  \atwoim   
\ \  {\text{for each}}\ \isan\  
 \ (i\in \Z), \nonum \\
\chanicI&:=0 \ \  {\text{for any   other}}\ \I\Subset\Z. 
\end{align}
Note  that  $\atwoip \atwoicr  \atwoim\in \Flisano$.
We see  that  $\chanic$ is $2$-periodic by the lattice translation 
on $\Z$  and that 
its  range is  finite  $r=2$.
 Therefore  $\chanic$  defined above 
is a bounded finite-range   assignment  of local fermion charges
on $\Z$, namely  $\chanic$ belongs to $\Chasp$.

We will  see that  the nilpotent condition 
\eqref{eq:nil-delcha}  necessary  for the 
  supersymmetry   is saturated  by $\delchanic$.
By noting the formula  \eqref{eq:deldelA-pair} of  
Proposition \eqref{prop:delcha-nil}
we   have to look for  only    
 the pairs  $\X_1\Subset\Z$ and $\X_2\Subset\Z$  
such that $\X_1\cap \X_2 \ne\emptyset$ and 
$\chanic(\X_2) \chanic(\X_1)\ne 0$. By the formula 
\eqref{eq:chanic} and  
the CARs \eqref{eq:CAR} we  verify that such pair does not exist.
 Thus  $\delchanic\cdot\delchanic(A)=0$
 holds  for every  $A\in \core$.  Accordingly  $\chanic\in \Chaspn$.

We will show  the existence of supersymmetric  states 
 with respect to  $\delchanic$.
Let  $\vp_{\rm{emp}}$ denote  
 the  unique state of  $\Fl$ determined by   
\begin{equation}
\label{eq:Fock}
\vp_{\rm{emp}}(\ajcr\aj)=0 \  {\text{for all}}\  j\in \Z.
\end{equation}
This  is a Fock state. 
Similarly one may  take  the fully occupied state $\vp_{\rm{occup}}$  
determined by 
\begin{equation}
\label{eq:occupied}
\vp_{\rm{occup}}(\aj\ajcr)=0\   {\text{for all}}\  j\in \Z.
\end{equation}
 For any   $A\in \core$ and $j\in \Z$,
\begin{align*}
0=\vp_{\rm{emp}}(A \atwojp \atwojcr  \atwojm)
=\vp_{\rm{emp}}( \atwojp \atwojcr  \atwojm A), \nonum\\  
0=\vp_{\rm{occup}}(A \atwojp \atwojcr  \atwojm)= 
\vp_{\rm{occup}}( \atwojp \atwojcr  \atwojm A),
\end{align*}
since   by the Cauchy-Schwarz inequality  
\begin{align*}
&\left| \vp_{\rm{emp}}(A \atwojp \atwojcr  \atwojm) \right|^{2}
\nonum\\
&\le \vp_{\rm{emp}}\left(\bigl(A \atwojp \atwojcr\bigr) 
 \bigl(A \atwojp \atwojcr\bigr)^{\ast} \right)
\cdot  \vp_{\rm{emp}}\bigl(\atwojmcr   \atwojm \bigr)
=0,
\end{align*} 
\begin{align*}
&\left| \vp_{\rm{emp}}(\atwojp \atwojcr  \atwojm A)\right|^{2}
=\left|\vp_{\rm{emp}}(-\atwojcr \atwojp \atwojm A)\right|^{2}
=\left|\vp_{\rm{emp}}(\atwojcr \atwojp \atwojm A)\right|^{2}
\nonum \\
&\le \vp_{\rm{emp}}\bigl(\atwojcr   \atwoj \bigr)
\cdot\vp_{\rm{emp}}\left(\bigl(\atwojp \atwojm A\bigr)^{\ast}
\bigl(\atwojp \atwojm A\bigr) \right)= 0,
\end{align*} 
and  similar  results hold for $\vp_{\rm{occup}}$.
Thus we obtain for any $A\in \core$ and $\X\Subset \Z$ 
\begin{align*}
0=\vp_{\rm{emp}}\bigl(A \chanic(\X)\bigr)
=\vp_{\rm{emp}}\bigl( \chanic(\X) A\bigr), \nonum\\  
0=\vp_{\rm{occup}}\bigl(A \chanic(\X)\bigr)= 
\vp_{\rm{occup}}\bigl( \chanic(\X) A\bigr).
\end{align*}
These yield  for any  $A\in \core$
 \begin{equation}
\label{eq:delFock}
\vp_{\rm{emp}}\bigl( \delchanic(A) \bigr)=
 \sum_{ \X \Subset \Z } \vp_{\rm{emp}} 
\bigl( [\chanic(\X), \;  A]_{\gamma} \bigr)=0,
 \end{equation}
and  
 \begin{equation}
\label{eq:delFock}
\vp_{\rm{occup}}\bigl( \delchanic(A) \bigr)=0.
 \end{equation}
Namely 
  both $\vp_{\rm{emp}}$ and $\vp_{\rm{occup}}$ are invariant 
under  $\delchanic$.  We have shown  $\chanic\in \Chaspnun$.
By  Theorem \eqref{thm:main}
 $\chanic\in \Chaspnun$ generates 
  supersymmetric $\cstar$-dynamics on the CAR algebra. 

In the following  we  provide  other two  examples. 
Let 
\begin{align}
\label{eq:chafen}
\chafeni&:= \ai  P_{<i>}    
\ \ {\text{for each}}\ \iryodonari\ 
 \ (i\in \Z), \nonum \\
\chafenI&:=0 \ \  {\text{for any  other}}\ \I\Subset\Z, 
\end{align}
where 
\begin{align}
\label{eq:Pi}
P_{<i>}:= (1- \aimcr  \aim) (1- \aipcr  \aip)    
\ \  {\text{for each}}\ \ i\in \Z.
\end{align}
This model  is due to   Fendly et al. \cite{FEN1}.
We easily  verify  that  $\chafen \in \Chaspn$ by 
 using  Proposition \eqref{prop:delcha-nil}.
We  also see  that the  state $\vp_{\rm{occup}}$ 
given in  \eqref{eq:occupied} is invariant under 
 the superderivation $\delchafen$ associated with $\chafen$. 
Hence  $\chafen\in \Chaspnun$.

One may consider  the following  much simpler   model. 
\begin{align}
\label{eq:chatriv}
\chatrivi&:= \ai      
\ \ {\text{for each}}\ i\in \Z, \nonum \\
\chatrivI&:=0 \ \  {\text{for any  other}}\ \I\Subset\Z. 
\end{align}
This is given by deleting  the projections $P_{<i>}$ 
in the form   \eqref{eq:chafen} of $\chafen$. 
We see   that  $\chatriv  \in \Chaspn$ by 
  Proposition \eqref{prop:delcha-nil}.
The corresponding time generator  s $\Derhchatriv\equiv
\delchatrivast\cdot\delchatriv+ \delchatriv \cdot\delchatrivast$
 is a   zero map on $\core$. 
So  the  time evolution $\{\alchatrivt\,;\;t\in \R\}$
on the total system $\Fl$ is  trivial. 
The supersymmetry  for 
 $\chatriv$ is spontaneously broken, since
for each $i\in\Z$
\begin{equation*}
 \delchatriv(\aicr)=\{ \ai, \;  \aicr\}=\unit,
 \end{equation*}
and then  for any  state $\ome$ on $\Fl$ 
\begin{equation*}
 \ome\left(\delchatriv(\aicr)\right) =\ome(\unit)=1\ne 0.
 \end{equation*}

\begin{rem}
\label{rem:FENunphysical}
The fully occupied state   $\vp_{\rm{occup}}$   
  may be   unphysical  
  for the Fendly's model, see  \cite{FEN1}.
However,  there  exist  (many)   ``physical''  supersymmetric 
 states  on  finite regions as noted in   
 \cite{FENchargefrust, vanEER}.
Any  cluster point  of such 
 physical states  in  the infinite-volume limit 
  gives  a   supersymmetric state by  Proposition \eqref{prop:infinite-volume}.
We can use  any such  supersymmetric  state  
in the place of $\vp_{\rm{occup}}$.
\end{rem}

\begin{rem}
\label{rem:brokentricial-case}
We have seen that 
 $\chatriv$ gives a  supersymmetry  breaking model.
Hence Proposition \eqref{prop:allclosable-unbroken} 
  can not  be applied to  $\chatriv$.
 Nevertheless,  
 the commutativity relations 
between its  (trivial) global   time evolution 
and its associated  superderivations 
 as in Proposition {\rm{\eqref{prop:koukan-lat}}}  are obviously saturated.
\end{rem} 

\section{Abstraction}
\label{sec:ABS}
One encounters  various  difficulties 
 to formulate  supersymmetry in $\cstar$-algebra.
 It seems, however, that those are   mixed up 
in the  literature. 
  We intend to  find an appropriate  solution to each of them.
 Buchholz-Grundling  have 
succeeded in formulating a simple supersymmetry model  
 by introducing  a new  $\cstar$-algebra called  the resolvent algebra
 in \cite{BUGR}.
 In this work  \cite{BUGR}
the crucial difficulty lies at unboundedness  of boson fields.
On the other hand,  
   we have focused on fermion lattice systems in this paper.
We have  discussed   $\cstar$-algebraic 
formulation of hidden  supersymmetry in fermion lattice 
 systems where  no  boson field exists.
  As the  fermion  system is     
 defined on  the CAR algebra,  we can make use of several    
  techniques of  $\cstar$-algebra theory. 
 We have seriously considered  some mathematical  problems
 related to the infinite volume limit.

In this final section, we shall provide a general scheme of supersymmetric 
 $\cstar$-dynamical systems based on our  $\cstar$-algebraic 
formulation of  supersymmetric fermion lattice systems.
 By  this abstraction  we can see  the general  structure  more clearly. 

Let $\Fl$ denote a general graded $\cstar$-algebra 
and let $\gamma$ denote  its grading 
 automorphism $\gamma$.
 Let  $\del$ be  a  superderivation of 
 $\Fl$ whose domain  $\core$ is a globally 
 $\gamma$-invariant $\ast$-subalgebra.
 Let  $\{\alt\,;\;t\in \R\}$ denote  
 a one parameter group of $\ast$-automorphisms of $\Fl$. 
The superderivation $\del$ and  its conjugate superderivation 
$\delast$ generate  supersymmetry transformation on $\Fl$, and 
 $\{\alt\,;\;t\in \R\}$ denotes a global time evolution on $\Fl$.
 The pair  $\{\del, \; \alt\}$ is  the   basic building kit. 
We  shall list  assumptions on $\{\del, \; \alt\}$ in the following.
\ \\

The first assumption  is  obvious.\\
$\bullet$ Time evolution preserves the grading:
\begin{equation}
\label{eq:alteven}
\alt \cdot \gamma=\gamma \cdot \alt \ \ {\text{for each}}\ t\in\R.
\end{equation}

We need some  continuity for  the time evolution with respect to time.\\
$\bullet$ Pointwise norm continuity: 
\begin{equation}
\label{eq:strongalt}
\lim_{t\to 0}\|\alt(F)-F\|\longrightarrow 0 
\ {\text{for every}}\ \ F \in \Fl.
\end{equation}
This  condition on  time evolution is 
 satisfied by quantum spin lattice models 
 and fermion lattice models for short-range interactions \cite{BR}.
 However, this can  not be expected for general 
  boson systems \cite{BR}. 
If the time evolution  $\{\alt\,;\;t\in \R\}$
satisfies  this  continuity condition, then  
 there exists  its infinitesimal generator defined by 
\begin{align}
\label{eq:derh-i}
\domderh&
:=\bigl\{X\in \Fl \,;\ \ \lim_{t\to 0} \frac{1}{t}\bigl( \alt(X)-X\bigr)
\ {\text{exists in norm in}}\ \Fl \bigr\},\notag\\
\Derh(X)
&:=-i\frac{d}{dt}\alt(X)\!\!\Bigm|_{t=0}\in\Fl \ \;{\text{for}}\ X\in \domderh.
\end{align} 
%%%%
 Of course, the Leibniz rule is satisfied by this: 
\begin{equation*}
\label{eq:leibniz}
\Derh(XY)=\Derh(X)Y+X \Derh(Y)\ \  {\text {for}}\ X, Y\in \domderh.
\end{equation*}
It has been   known that     
$\domderh$  is a  norm-dense  $\ast$-subalgebra 
of $\Fl$, see \cite[Proposition 3.1.6]{BR}.  
From   \eqref{eq:alteven}  
 it follows that 
\begin{equation*}
 \gamma(\domderh)=\domderh,\quad 
\Derh\cdot \gamma=\gamma \cdot \Derh\ \ 
{\text {on}}\ \domderh.
\end{equation*}
\begin{rem}
\label{rem:bar}
  As defined  in \eqref{eq:derh-i} the derivation $\Derh$ 
  is  {{closed}},  whereas  the superderivation  
$\del$  on $\core$ is {{not}}  assumed to be closed. 
Later $\del$ will be assumed to be {\it{closable}}.
\end{rem}

To  encode  essential information of dynamics on 
 the domain of the superderivation we postulate the following.\\
$\bullet$ The  domain of the  superderivation is ``large":  
\begin{equation}
\label{eq:coredense}
\core\ {\text{is  norm-dense in }}\ \Fl.
\end{equation}

To  relate 
 the superderivation with  the time evolution
 the following assumption will be  convenient.\\ 
$\bullet$ The domain of  the superderivation is included 
in that of the time generator:
\begin{equation}
\label{eq:Derhcore}
\core\subset \domderh.
\end{equation}

We assume 
   differentiability  of the  superderivation in the following sense.
 This is  crucial.\\
$\bullet$ Differentiability of the superderivation:  
\begin{equation}
\label{eq:DIFFalg}
\del(\core)\subset \core.
\end{equation}
It is immediate to see that  
 the associated  superderivations 
$\delast$, $\delsf$ and $\delss$   satisfy  
  this   differentiability as well. 
 Actually from \eqref{eq:delastDEF} \eqref{eq:DIFFalg}
and the  $\gamma$-invariance of $\core$,
 it follows that    
\begin{equation}
\label{eq:diffalgast}
\delast(\core)\subset \core. 
\end{equation}
From  \eqref{eq:dels-futatu} \eqref{eq:DIFFalg}
\eqref{eq:diffalgast} it follows that  
\begin{equation}
\label{eq:diffalgsym}
 \delsf(\core)\subset \core,\quad \delss(\core) \subset \core.  
\end{equation}
By \eqref{eq:DIFFalg}  \eqref{eq:diffalgast} \eqref{eq:diffalgsym}  
  any composition of $\del$, $\delast$, $\delsf$, $\delss$ can be  
defined on $\core$.

\begin{rem}
\label{rem:QALG}
The  differentiability of superderivations is due to the 
 differential  graded algebra (DGA), see \cite{MANIN}. 
 It is  one of  the desiderata  of  
`the quantum algebra' by Jaffe et al. \cite{JAFLO, JAFLOK}.
\end{rem}

We will list  more involved  assumptions.\\
$\bullet$ The domain of the superderivation is  a core for the time generator:  \begin{equation}
\label{eq:corecore-Derh}
\overline{\Derh|_{\core}}=\Derh,
\end{equation}
where the bar on
$\Derh|_{\core}$
denotes the norm closure. 

 We  assume the following   topological property  on the superderivations. \\ 
$\bullet$ Norm closability of the superderivations:
\begin{equation}
\label{eq:normcloable}
\del: \core \mapsto\Fl   \ \; {\text{is norm-closable}},
\end{equation}
and 
\begin{equation}
\label{eq:normclosable-s}
\delsf: \core \mapsto\Fl  \ \; {\text{is norm-closable}},\quad 
\delss: \core \mapsto\Fl  \ \; {\text{is norm-closable}}. 
\end{equation}
It has been noted in 
 Proposition \eqref{prop:ast-normclo}
 that  \eqref{eq:normcloable} implies that 
\begin{equation}
\label{eq:normclosable-ast}
\delast: \core \mapsto\Fl   \ \; {\text{is norm-closable}}.
\end{equation}
We denote the norm closure of  $\del$ by $\delcl$ and its extended domain by 
$\domdelcl$. 
\begin{rem}
\label{rem:allnorm-closable}
For the  supersymmetric  fermion lattice systems
presented  in  Sect.\ref{sec:SUSYfer}, 
 we  have guaranteed   all of  
\eqref{eq:normcloable} \eqref{eq:normclosable-ast}
\eqref{eq:normclosable-s}
in Propositions \eqref{prop:allclosable-unbroken}. 
\end{rem}

When the superderivations are   norm-closable,  
we can  {{assume}} the following    relations.\\ 
$\bullet$ Time evolution preserves  supersymmetry:
\begin{align}
\label{eq:koukandel}
&\altcore \subset \domdelcl\ \ {\text{and}} \ \   
\delcl \cdot \alt=\alt \cdot \del
{\text{\ \ on\ \ }} \core \ \ {\text{for  every}}\  t\in \R,
\nonum\\
&\altcore \subset \domdelastcl\ \ {\text{and}} \ \   
\delastcl \cdot \alt=\alt \cdot \delast
{\text{\ \ on\ \ }} \core \ \ {\text{for  every}}\  t\in \R,\nonum\\
&\altcore \subset \domdelsfcl\ \  {\text{and}} \ \   
\delsfcl \cdot \alt=\alt \cdot \delsf {\text{\ \ on\ \ }} \core
\ \ {\text{for  every}}\  t\in \R, \nonum\\
&\altcore \subset \domdelsscl \ \ {\text{and}} \ \   
\delsscl \cdot \alt=\alt \cdot \delss
{\text{\ \ on\ \ }} \core \ \ {\text{for  every}}\  t\in \R.
\end{align}

\begin{rem}
\label{rem:KAKAN-assumption}
The assumption 
  \eqref{eq:koukandel}   seems   indispensable 
 for any natural  supersymmetry theory, since
 the commutativity  between  
the supercharges  and the supersymmetric Hamiltonian  follows 
 directly from the supersymmetry algebra  
 as  noted  in Sect.\ref{sec:overview}.
On the other hand,  if  the 
  domain $\core$ of the superderivation  is not  invariant 
 under the global time evolution, then there requires   subtle care. 
We have considered this non-trivial problem   in 
 Proposition \eqref{prop:koukan-lat}   in Sect.\ref{sec:SUSYfer}
 for the case of fermion lattice systems.
\end{rem}

With all the assumptions given so far,  
we shall propose  supersymmetric $\cstar$-dynamics.
\begin{defn}
\label{defn:susydyn-abs}
Let $\del$ denote  a  superderivation 
of a graded $\cstar$-algebra $\Fl$ and let $\core$
 denote its   domain subalgebra. 
Let $\{\alt\,;\;t\in \R\}$ denote a  
 strongly continuous  one-parameter
group of $\ast$-automorphisms  of $\Fl$ and 
 let  $\Derh$ denote its   generator.
All the assumptions  stated  in this section are satisfied.
If $\del$ is  nilpotent, i.e.  
\begin{equation}
\label{eq:nilp-gen}
\del\cdot\del=\zeromap
{\text {\ \ on\ \ }}\core,  
\end{equation} 
and the following  relation holds 
\begin{equation}
\label{eq:susy-gen}
\Derh=\delast\cdot\del+\del\cdot\delast {\text {\ \ on\ \ }}\core,
\end{equation}
then it is said that $\{\del,\;\alt\}$ 
  generates  a supersymmetric $\cstar$-dynamics.
\end{defn}

As we  have seen  in 
 Lemma \eqref{lem:ast-nil} in  Sect.\ref{sec:SUSYfer},
  the  superderivation $\del$ is nilpotent if and only if 
   $\delast$ is  nilpotent:
\begin{equation}
\label{eq:nilpast-gen}
\delast\cdot\delast=\zeromap  
{\text {\ \ on\ \ }} \core. 
\end{equation}

We note the following general remark on superderivations.
\begin{prop}
\label{prop:hermite-nil-Fuka}
Suppose that  a superderivation $\del$ is nilpotent
and hermite.
 Then the supersymmetric dynamics induced by $\{\del,\;\alt\}$ 
 is  trivial. 
\end{prop}
\begin{proof}
 By \eqref{eq:nilp-gen}
 \eqref{eq:susy-gen} 
\eqref{eq:nilpast-gen} together with the 
 hermite property   $\delast=\del$, we obtain 
 $\Derh=\zeromap$  on $\core$.
By this and   \eqref{eq:corecore-Derh}
 the   strongly continuous time evolution 
  $\{\alt\,;\;t\in \R\}$ is trivial, i.e. $\alt$ is an identity map
 for each $t\in \R$.
\end{proof}

We can rewrite  the  supersymmetric $\cstar$-dynamics  
given above in terms of the  hermite superderivations.
\begin{prop}
The set of relations \eqref{eq:nilp-gen} and 
\eqref{eq:susy-gen}  in 
   Definition \eqref{defn:susydyn-abs} is equivalent 
to the combination of  
\begin{equation}
\label{eq:nilsymdel}
\delsf\cdot\delss+\delss\cdot \delsf=\zeromap
{\text {\ \ on\ \ }}\core,  
\end{equation}
and 
\begin{equation}
\label{eq:susysym}
\Derh={\delsf}^{\!\!\!2}={\delss}^{\!\!\!2}{\text {\ \ on\ \ }}\core,  
\end{equation}
 where the hermite superderivations   $\delsf$ and $\delss$ 
are given in \eqref{eq:dels-futatu}.
\end{prop}
\begin{proof}
The proof has been  essentially done in Proposition \eqref{prop:susy-cha-sym}
 in  Sect.\ref{sec:SUSYfer}.
\end{proof}

\begin{rem} 
\label{rem:analyticJAFLO}
 Another   $\cstar$-algebraic framework of  supersymmetry   
 is proposed  by Jaffe et al. in \cite[Sect.2]{JAFLOK}.
 Here the superderivation $\del$ of a graded $\cstar$-algebra $\Fl$ 
is  defined on  the analytic algebra $\Fl_{\al}$
 that consists of  the  {\it{whole entire analytic elements}} for the given 
 strongly continuous time automorphism group $\alt$, 
 and the supersymmetry relation  is defined  on  $\Fl_{\al}$.
 Its justification is not given there. 
 Theorem \eqref{thm:main} in Sect.\ref{sec:SUSYfer} gives  
 a concrete realization of 
 $\cstar$-algebraic supersymmetry  on the  fermion lattice system   
  taking  the local algebra for  the domain of  superderivations.
    This  norm-dense subalgebra  
  is {\it{strictly}} contained in  $\Fl_{\al}$.
\end{rem}

Finally, we shall propose  a naive  question.  
How should the listed assumptions   be changed  
 to deal with  supersymmetry
  between  fermions and bosons?
 By comparing this work with  
   \cite{BUGR}  \cite{MORIYAnew} we shall  give some comments.
First the norm topology 
used  in \eqref{eq:strongalt} \eqref{eq:coredense} 
 should be  replaced by some  weaker one. 
Second, the differentiability of the superderivation
 \eqref{eq:DIFFalg}  is likely  not  to be satisfied.
 So it should be removed or altered. 
Third,  the assumption 
  \eqref{eq:koukandel}   may  be trivial 
 for quantum field theory 
if the domain of superderivations 
 is made by  local subsystems only. Note  that    
general  time evolution in fermion (quantum spin) lattice systems
 is non-local, 
 whereas  it is  local  in  relativistic systems
 due to  the  finite velocity of its propagation.

\subsection*{Acknowledgment}
I thank  a referee of AHP  for several remarks that improve the proof 
 of the main result. 
I thank  Asao Arai, Izumi Tsutsui and Yu Nakayama    
 for useful discussion   on SUSY quantum mechanics.
 I thank  Detlev Buchholz     
 for discussion  on $\cstar$-algebraic approach to SUSY.
This work is partly supported by JSPS-kakenhi 21740128.
\end{document}